\RequirePackage{fix-cm}

\documentclass[twocolumn,epjc3]{svjour3}  

\smartqed  

\RequirePackage{graphicx}
\RequirePackage{hyperref}
\RequirePackage{epsfig,graphics,colordvi}
\RequirePackage{amsmath}
\RequirePackage{amssymb}
\RequirePackage{mciteplus}
\RequirePackage{dsfont}
\RequirePackage[english]{babel}
\RequirePackage{caption}
\RequirePackage{subcaption}
\RequirePackage{sidecap}
\RequirePackage{multirow}
\RequirePackage{dsfont}

\newcommand{\tr}{\operatorname{tr}}
\newcommand{\te}{\text}
\newcommand{\nn}{\nonumber}
\newcommand{\cd}{\cdot}

\newcommand{\eps}{\varepsilon}
\newcommand{\etapr}{\eta\hspace{0.05em}'}
\newcommand{\dint}{\text{d}}
\newcommand{\fpi}{f_{\hspace{-0.1em} \pi}}

\journalname{Eur. Phys. J. C}
\begin{document}

\title{Reactions with pions and vector mesons in the sector of odd intrinsic parity}


\author{C.\ Terschl\"usen\thanksref{eM,inst1} \and B.\ Strandberg\thanksref{inst1} \and S.\ Leupold\thanksref{inst1} \and F. Eichst\"adt\thanksref{inst2}}  

\thankstext{eM}{e-mail: carla.terschluesen@physics.uu.se}

\institute{Institutionen f\"or fysik och astronomi, Uppsala Universitet, Box 516, 75120 Uppsala, Sweden \label{inst1}
           \and
           Institut f\"ur Theoretische Physik, Justus-Liebig-Universit\"at Gie\ss{}en, Heinrich-Buff-Ring 16, 35392 Gie\ss{}en, Germany \label{inst2}}

\date{Received: date / Accepted: date}

\maketitle

\begin{abstract}
 The Wess-Zumino-Witten structure is supplemented by a simple vector-meson Lagrangian where the vector mesons are described by antisymmetric tensor fields. With the $\rho$-$\omega$-$\pi$ coupling as the only parameter in the sector of odd intrinsic parity, i.e.\ without additional contact terms, one can achieve a proper description of the decay of an $\omega$-meson into three pions, the single- and double-virtual pion transition form factor and the three-pion production in electron-positron collisions. 
\end{abstract}

\section{Introduction} \label{sec:ind}
One of the challenges of particle physics at its low-energy frontier is the investigation of the properties of hadrons and their interactions. This is important for a better understanding of quantum chromodynamics (QCD) in its non-perturbative low-energy regime and has also numerous implications for the study of hot and dense strongly interacting matter and the phase structure of QCD \cite{Friman:2011zz}. In addition, for searches of physics beyond the standard model the hadronic (standard-model) contributions to a given observable need to be under control so that one can compare data with the respective standard-model prediction. For instance, the processes studied in the present work, the pion transition form factor and the hadron production in electron-positron collisions, constitute important input for the gyromagnetic ratio of the muon \cite{Jegerlehner:2009ry, Czerwinski:2012ry}. 

QCD is equivalent to chiral perturbation theory ($\chi$PT) 
\cite{Weinberg:1978kz,Gasser:1983yg,Gasser:1984gg,Scherer:2002tk} in the regime of very low energies where no other hadrons are excited besides the quasi-Goldstone bosons of spontaneously broken chiral symmetry. In the sector of even intrinsic parity\footnote{For a meson of spin $J$ and parity $P$ the intrinsic parity is given by $(-1)^{P+J}$. In particular, scalars and vectors have positive (even) intrinsic parity while it is negative (odd) for pseudoscalars and pseudovectors (axialvectors). Interactions in the sector of odd intrinsic parity contain the Levi-Civita tensor while interactions in the sector of even intrinsic parity do not.} the leading-order Lagrangian is given by 
\begin{align}
\mathcal{L_{\chi \rm PT}} = \fpi^2 \, {\rm tr}(U^\dagger_\mu \, U^\mu) 
  + \frac12 \, \fpi^2 \, {\rm tr}(\chi_+)  
\label{eq:LChPT}
\end{align}
with
\begin{align}
  &U_\mu  :=  \frac12 \, u^\dagger \, (D_\mu U) \, u^\dagger = - U_\mu^\dagger \, , \nn \\
  &D_\mu U := \partial_\mu U - i r_\mu \, U + i \, U \, l_\mu  \,, \nn \\
  &U(\Phi)  =   \exp\left(+i \, \Phi/\fpi\right), \ U =: u^2 \,, \nn \\
  &\chi_+  :=  \frac12 \, \left(u \chi^\dagger u + u^\dagger \chi\, u^\dagger \right)  \,.    
  \label{eq:defderextsourcecomb2}
\end{align}
Hereby, $r_\mu := v_\mu + a_\mu$, $l_\mu := v_\mu - a_\mu$ and $\chi := 2 \, B_0 \, (s+ip)$ including the external vector, axialvector, scalar and pseudoscalar sources $v_\mu$, $a_\mu$, $s$ and $p$, respectively. The pion fields are collected in the matrix
\begin{align}
 \Phi = \begin{pmatrix}
         \pi^0 & \sqrt{2} \, \pi^+ \\ \sqrt{2} \, \pi^- & -\pi^0
        \end{pmatrix}.
\end{align}
If the external sources are switched off, one finds in the two-flavor sector
\begin{align}
  \label{eq:defchi0}
  \chi \mapsto \chi_0 := 2\,B_0 \, \mathcal{M} := 2\,B_0 \, \text{diag}\left(m_u, \, m_d \right) \approx m_\pi^2 \, \mathds{1}_{2 \times 2}\,, 
\end{align}
where isospin breaking has been neglected. Electromagnetism is introduced by replacing $v_\mu \mapsto -e \,Q \mathcal{A}_\mu$ with the positron charge $e$, the photon field $\mathcal{A}_\mu$ and the two-flavor quark-charge matrix
\begin{align}
 Q := \text{diag} \left( \frac{2}{3}, \, - \, \frac{1}{3} \right).
\end{align}

In the sector of odd intrinsic parity ($\varepsilon$ sector) the leading-order action is governed by the chiral anomaly, which gives rise to the Wess-Zumino-Witten (WZW) action \cite{Wess:1971yu,Witten:1983tw}. For two flavors and restricted to an external electromagnetic field, one finds the WZW Lagrangian (see, e.g., \cite{Scherer:2002tk, Kaiser:2000ck, Bar:2001qk} for the complete $SU(N)$ structure)
\begin{align}
 &\mathcal{L}_{\rm WZW} \nn \\
 &= - \,i \, \frac{n \, e^2}{48 \pi^2} \, \eps^{\mu\nu\alpha\beta} \, \tr\left( Q \left( \partial_\mu U \, U^\dagger + U^\dagger \, \partial_\mu U \right) \right) \partial_\nu \mathcal{A_\alpha} \, \mathcal{A}_\beta \nn \\
 &\phantom{=}{}\ + \,\frac{n \, e^{\phantom{2}}}{144 \pi^2}\, \eps^{\mu\nu\alpha\beta} \, \tr\left( \partial_\mu U \, U^\dagger \, \partial_\nu U \, U^\dagger \, \partial_\alpha U \, U^\dagger \right) \mathcal{A}_{\beta} \,. \label{eq:LWZW}
\end{align}
As shown by Witten \cite{Witten:1983tw}, the modulus of $n$ is equal to the number of quark colors, $N_c$.

Formally the leading-order Lagrangian (\ref{eq:LChPT}) for even intrinsic parity is of order $p^2$ while the WZW action is of order $p^4$, where $p$ denotes a typical momentum in the order of a mass of a quasi-Goldstone boson. To improve the accuracy of a $\chi$PT calculation or to describe reactions at somewhat larger energies one needs the respective Lagrangians of next-to-leading order. For the sector of even intrinsic parity the $p^4$ Lagrangian has been worked out in \cite{Gasser:1983yg} for two and in \cite{Gasser:1984gg} for three light flavors. For the $p^6$ Lagrangian in the sector of odd intrinsic parity see \cite{Ebertshauser:2001nj} and references therein. It also has been shown already in \cite{Gasser:1983yg} that the low-energy constants at order $p^4$ are essentially saturated by vector-meson exchange for all channels where vector mesons can contribute\footnote{For extensions to three flavors (including also other resonances) see \cite{Ecker:1988te,Ecker:1989yg} and to the sector of odd intrinsic parity see \cite{Kampf:2011ty}.}. This implies that $\chi$PT at next-to-leading order is equivalent to the following Lagrangian, which couples the Goldstone bosons and external fields to vector mesons represented by antisymmetric tensor fields \cite{Ecker:1988te,Ecker:1989yg}:
\begin{align}
 \mathcal{L}_{\rm vec} =& \,\frac{i}{2} \, f_V \, h_P \, \tr\left( U_\mu \,\Phi^{\mu\nu} \,U_\nu \right) + \, \frac{1}{2}\, f_V \tr\left( \Phi^{\mu\nu} f_{\mu\nu}^+ \right) \nn \\
 &{} - \, \frac{1}{4} \tr \left( \left( D^\mu \Phi_{\mu\alpha} \right) \left( D_\nu \Phi^{\nu\alpha} \right) \right) \, + \, \frac{1}{8}\, m_V^2 \, \tr \left( \Phi^{\mu\nu} \Phi_{\mu\nu} \right) 
\label{eq:Lvec}
\end{align}
with the two-flavor vector-meson matrix
\begin{align}
 \Phi_{\mu\nu} = \begin{pmatrix} 
                   \rho_{\mu\nu} + \omega_{\mu\nu} & \sqrt{2} \, \rho^+_{\mu\nu} \\
                   \sqrt{2} \, \rho^-_{\mu\nu} & -\rho_{\mu\nu} + \omega_{\mu\nu}         
                 \end{pmatrix}
\end{align}
and the building block
\begin{align}
 f_{\mu\nu}^+ = \frac{1}{2} &\left\{ u \left( \partial_\mu l_\nu - \partial_\nu l_\mu - \, i\left[ l_\mu, l_\nu \right] \right) u^\dagger \right. \nn \\
 &\phantom{\left\{\right.}{}\left. + \, u^\dagger \left( \partial_\mu r_\nu - \partial_\nu r_\mu -  i\left[ r_\mu, r_\nu \right] \right) u \right\} .
\end{align}
The covariant derivate $D^\mu$ acts on the vector-meson field as
\begin{align}
  &D_\mu \Phi^{\mu\nu} = \partial_\mu \Phi^{\mu\nu} + \left[\Gamma_\mu, \Phi^{\mu\nu} \right], \nn \\
  &\Gamma_\mu = \,\frac{1}{2} \left\{ u^\dagger \left(\partial_\mu -  i\, r_\mu \right) u \,+\, u \left( \partial_\mu - i\, l_\mu \right) u^\dagger \right\}.
\end{align}
The mass $m_V$ in (\ref{eq:Lvec}) denotes the (common) mass of $\rho$- and $\omega$-mesons. 

Note that interactions with other resonances, as introduced in \cite{Ecker:1988te}, are ignored since they are not important for our purposes. We also note that we use the conventions of \cite{Terschlusen:2012xw} in the present work. The coupling constants $f_V$ and $h_P$ are related to $F_V$ and $G_V$ used in \cite{Ecker:1988te,Ecker:1989yg} by $F_V = f_V$ and $G_V = h_P \, f_V /4$.
For the sector of odd intrinsic parity we supplement (\ref{eq:Lvec}) by 
\begin{align}
 \mathcal{L}_{A} = \,\frac{i}{8}\, h_A \, \eps^{\mu\nu\alpha\beta} \, \tr\left( \left\{\Phi_{\mu\nu}, D^\tau \Phi_{\tau\alpha} \right\} U_\beta \right) .
\label{eq:LA}
\end{align}
This term, which induces a $\rho$-$\omega$-$\pi$ interaction, resembles the well-known $g_A$ term of pion-nucleon scattering \cite{Scherer:2002tk}. The $g_A$ term is the simplest interaction term of two nucleons and a pion compatible with chiral symmetry. In baryon $\chi$PT it is of leading order. 
Correspondingly, the $h_A$ term (\ref{eq:LA}) is the simplest interaction term of two vector mesons and one pion compatible with chiral symmetry. 
In the following we will use the presented interactions of \eqref{eq:LChPT}, \eqref{eq:LWZW}, \eqref{eq:Lvec} and \eqref{eq:LA} outside the realm of $\chi$PT for energies where the vector mesons are active degrees of freedom. In that sense it is a phenomenological application of Lagrangians which have proven to be important in the low-energy regime of $\chi$PT.

The point we want to make in the present work is that the vector-meson Lagrangian given in (\ref{eq:Lvec}) and (\ref{eq:LA}) with its only three coupling constants provides a very economic way to successfully describe the low-energy interactions of pions, $\rho$- and $\omega$-mesons. Especially in the sector of odd intrinsic parity there is only the term (\ref{eq:LA}) on top of the WZW structure \eqref{eq:LWZW}. In particular this implies that the four-point interaction of one vector meson, $V$, with three pions is solely described as a two-step process of $2V$-$\pi$ ($h_A$ term) and $V$-$2\pi$ ($h_P$ term) \cite{Leupold:2008bp}. Correspondingly, the electromagnetic transition of $V$ to $\pi$ is solely described by $2V$-$\pi$ ($h_A$ term) and $V$-$\gamma$ ($f_V$ term) \cite{Terschluesen:2010ik}. In that sense our Lagrangian seems to support the concept of vector-meson dominance (VMD) \cite{Sakurai:1969}. On the other hand, we will see in the following that the reactions of pions with electromagnetism also include the respective lowest order $\chi$PT structures in addition to vector-meson terms. Thus, we do not find vector-meson dominance in its strict form and there are even cases where our approach suggests large deviations \cite{Terschluesen:2010ik,Leupold:2012qn}. Even more important (and not unrelated \cite{Ecker:1989yg}), however, is the fact that in our approach the vector mesons are represented by antisymmetric tensor fields and not, as traditionally done, by vector fields \cite{Sakurai:1969, Bando:1987br, Meissner:1987ge, Birse:1996hd}. Indeed, if one uses vector fields to represent the vector mesons, one needs significantly more terms to describe the interactions of pions and vector mesons in the sector of odd intrinsic parity \cite{Bando:1987br,Meissner:1987ge,Klingl:1996by,Harada:2003jx,Benayoun:2009im}. 
For instance, in the hidden-gauge formalism \cite{Bando:1987br,Harada:2003jx} four independent interaction terms (describing contact interactions of type $2V$-$\pi$, $V$-$3\pi$, $V$-$\gamma^{(*)}$-$\pi$ and $3V$-$\pi$) are introduced together with the WZW action. Three of them would be relevant for the processes discussed in the present work. Also in \cite{Klingl:1996by} three independent interaction terms are needed to describe $V$-$3\pi$ and $V$-$\gamma^{(*)}$-$\pi$ interactions using vector fields to represent the vector mesons. In contrast, as we will demonstrate in the following, to get a good description of the $\pi$-$\rho$-$\omega$ interactions, only the single interaction term (\ref{eq:LA}) has to supplement the WZW anomaly structure \eqref{eq:LWZW} in the sector of odd intrinsic parity.

Of course, the vector-meson Lagrangian (\ref{eq:Lvec}) and (\ref{eq:LA}), given in the antisymmetric tensor representation, can be rewritten into any vector representation along the lines described in \cite{Ecker:1989yg, Birse:1996hd, Bijnens:1995ii, Borasoy:1995ds, Leupold:2006bp}. In that way one would recover the structures of \cite{Klingl:1996by,Harada:2003jx,Benayoun:2009im}. But now the coupling constants of these structures would not be independent but interrelated, since they would be expressed in terms of $h_A$ (together with the $V$-$\gamma$ and $V$-$2\pi$ coupling constants which are present in any
approach). This reiterates our statement that the approach presented here is more economic in the description of the low-energy reactions of pions and vector mesons in the sector of odd intrinsic parity. 

The publication is organized in the following way: In the next section the coupling constants in the sector of even intrinsic parity are determined by considering the pion form factor and the decay of an $\omega$-meson into a dilepton. The reactions $\omega \to \pi^0 \gamma^{(\ast)}$ and $\omega \rightarrow 3\pi$ are discussed in section \ref{sec:omega decays} including the determination of the coupling constant in the sector of odd intrinsic parity. In the following section the single- and double-virtual pion transition form factor as well as the decays of a pion into one or two real photons and into two dielectrons are calculated. Thereafter, the scattering reaction $e^+ e^- \to 3\pi$ is considered (section \ref{sec:scattering}). In section \ref{sec:fudi} we discuss whether a deeper foundation of our Lagrangian is conceivable in the framework of an effective field theory. The last section provides a summary.
\section{Determination of the coupling constants in the sector of even intrinsic parity} \label{sec:coupling constants}
In the following we use the physical pion decay constant $\fpi = 92.2 \,\te{MeV}$. For $m_\pi$ we use the isospin averaged mass $(2 m_{\pi^+} + m_{\pi^0})/3 \approx 138\,\te{MeV}$ except where explicitly stated otherwise. In principle, the coupling constants $f_V$ and $h_P$ can be determined from decays of $\rho$- and $\omega$-mesons. However, since the $\rho$-meson is a broad resonance we prefer to use the pion form factor directly. Thereby, we proceed along the lines described in \cite{Leupold:2009} where the pion form factor is calculated including pion-pion rescattering. The main results from \cite{Leupold:2009} will be recalled here. 

\begin{figure}[h]
 \begin{center} $
 \begin{array}{ccccc}
 \includegraphics[trim = 0 26 0 -26, width=0.06\textwidth]{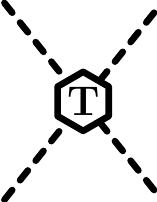} \vspace{1em}& 
 \large{\ = \ } &
 \includegraphics[trim = 0 26 0 -26, width=0.06\textwidth]{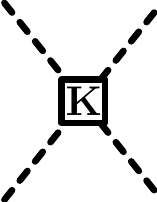} & 
 \large{\ + \ } &
 \includegraphics[trim = 0 26 0 -26, width=0.12\textwidth]{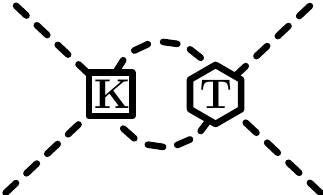} \\ 
 \includegraphics[trim = 0 26 0 -26, width=0.06\textwidth]{Bethe-Salpeter-2.pdf} \vspace{2em}& 
 \large{\ = \ } &
 \includegraphics[trim = 0 26 0 -26, width=0.06\textwidth]{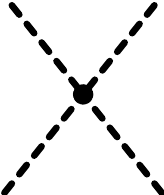} &
 \large{\ + \ } &
 \includegraphics[trim = 0 26 0 -26, width=0.09\textwidth]{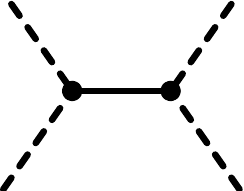} 
 \end{array} $
 \caption{Bethe-Salpeter equation (\textbf{top}) and its scattering kernel (\textbf{bottom}) for pion-pion scattering. The dashed lines represent the pions, the solid line the $\rho$-meson.}
 \label{fig:scatt-kernel}
 \end{center}
\end{figure}

In the center-of-mass system, the scattering amplitude for elastic pion-pion scattering can be decomposed into amplitudes $t_l$ with angular momentum $l$, 
\begin{align}
 \mathcal{M}_{\pi\pi \rightarrow \pi\pi} (s,\cos\theta) = \sum_{l} (2l+1)\, t_l(s) \, P_l(\cos\theta),
\end{align}
including the Legendre polynomials $P_l$. Since the process of interest for the pion form factor, $e^+ e^- \rightarrow \pi^+ \pi^-$, proceeds via a virtual photon in the $s$ channel, the orbital angular momentum of the two-pion system is fixed to $l = 1$. Thus, for an overall symmetric two-pion state the isospin has to be $I = 1$. Thereby, the scattering amplitude consists of two-particle reducible and irreducible parts. The irreducible parts give rise to the kernel of the Bethe-Salpeter equation which automatically produces the reducible ones. First, one approximates the scattering kernel $k_{\pi \pi \rightarrow \pi \pi}$ by a direct scattering described by $\chi$PT \eqref{eq:LChPT} and scattering via a $\rho$-meson arising from the $h_P$ term in \eqref{eq:Lvec} (see bottom panel of Fig. \ref{fig:scatt-kernel}),
\begin{align}
 k_{\pi \pi \rightarrow \pi \pi} &= k_{\,l=1, I=1}(s) \nn \\
 &=\frac{2}{3 \fpi^2} \, p_{\rm c.m.}^2(s) \left(1 - \, \frac{h_P^2 \, f_V^2}{8 \fpi^2} \, \frac{s}{s - m_{\rho, \rm bare}^2} \right)  \label{eq:kpipi}                               
\end{align}
including the center-of-mass momentum\newline $p_{\rm c.m.}(s) = \frac{1}{2} \sqrt{s - 4 \, m_{\pi^+}^2 }$.
Therewith, the Bethe-Salpeter equation can be calculated (see top panel of Fig. \ref{fig:scatt-kernel}),
\begin{align}
 t_{l=1, I=1}^{-1}(s) = k_{l=1, I=1}^{-1}(s) - I(s) = \frac{1-k_{l=1, I=1}(s)\, I(s)}{k_{l=1, I=1}(s)} \,.
\end{align}
Here, $I(s)$ denotes the loop function regularized by dimensional regularization with dimension $d= 4+ 2 \,\eps$, 
\begin{align}
 &I(s = p^2) \nn \\ &= - \, i\int \frac{\dint^d q}{(2\pi)^d} \, \frac{1}{\left[q^2 -m_\pi^2 + \, i\eta \right] \left[(q-p)^2 - m_\pi^2 + \, i\eta\right]} \,.
\end{align}
Renormalization is ensured via the replacement 
\begin{align}
 &I(s) \mapsto I(s) - \te{Re}\,I(\mu^2) = J(s) - \te{Re}\,J(\mu^2), \, \nn \\
 &J(s) = \frac{1}{16\pi^2} \left( 2 + \sigma(s) \, \log \, \frac{\sigma(s) - 1}{\sigma(s) +1} \right) \label{eq:def J}
\end{align}
with the phase space $\sigma(s) = 2 \,p_{\rm c.m.}(s) / \sqrt{s}$ and a new parameter, the renormalization point $\mu$. For simplicity, we choose $\mu = m_\rho = m_{\rho, \te{bare}}$ with the physical mass $m_\rho$ of the $\rho$-meson. Changes in $\mu$ can be traded essentially by changes in $m_{\rho, 
\te{bare}}$ \cite{Leupold:2009}.

\begin{figure}[h]
 \centering
 \includegraphics[trim = 40 50 0 60, width = 0.5\textwidth]{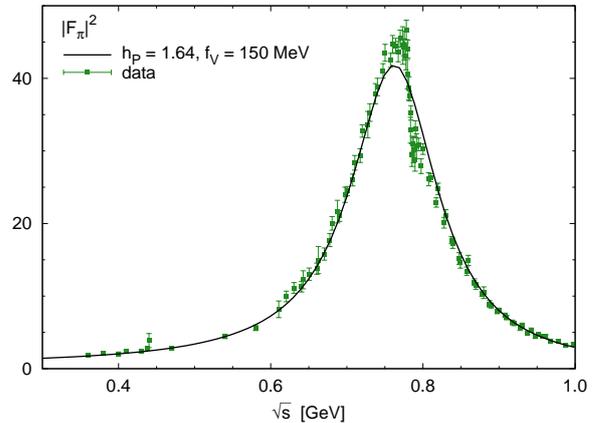}
 \caption{Modulus square of the pion form factor compared to data \cite{Barkov:1985ac, Akhmetshin:2001ig, Akhmetshin:2003zn}.}
 \label{fig:pion form factor}
\end{figure}

The pion form factor can now be expressed as
\begin{align}
 F_\pi(s) &= F_\pi^{\rm tree}(s) \left[ 1+ \left(J(s) - \te{Re}J(\mu^2) \right) t_{I=1, l=1}(s) \right] \nn \\
 &= \frac{F_\pi^{\rm tree}(s)}{1 - \left(J(s) - \te{Re}J(\mu^2) \right) k_{I=1, l=1}(s)}
\end{align}
with the tree-level pion form factor 
\begin{align}
 F_\pi^{\rm tree}(s) = 1 - \, \frac{h_P \, f_V^2}{4 \fpi^2} \, \frac{s}{s- m_{\rho, \te{bare}}^2}\,. \label{eq:Ftree pi}
\end{align}
By fitting the modulus square of the pion form factor to experimental data \cite{Barkov:1985ac, Akhmetshin:2001ig, Akhmetshin:2003zn} (see Fig. \ref{fig:pion form factor})\footnote{The additional narrow peak slightly to the right of the main peak in the data is caused by the appearance of the $\omega$-meson due to isospin violating $\rho$-$\omega$ mixing. Since our calculations are done in the isospin limit, we cannot describe this additional peak structure.}, one can essentially fix the parameter combinations
\begin{align}
 \frac{h_P f_V}{\fpi^2} = 29 \,\te{GeV}^{-1}, \ f_V =: f_V^\rho = 150 \, \te{MeV}. \label{param:fV}
\end{align}
This results in\footnote{The parameter $h_P$ was redefined in \cite{Terschlusen:2012xw}. For the old definition of $h_P$, we would have gotten $|h_P| = 0.32$ differing by less than $10\%$ from the values used in \cite{Leupold:2008bp, Leupold:2009}.}
\begin{align}
 h_P = 1.64\,. \label{param:hp}
\end{align}
Thereby, the sign of $f_V$ is pure convention in accordance with the definition that the vector-meson fields $\Phi_{\mu\nu}$ instead of $(-\Phi_{\mu\nu})$ produce vector mesons. After fixing the sign of $f_V$, there is no freedom left for sign choices in other interaction terms with an odd number of vector-meson fields. 

Alternatively, one can use the decay $\omega \rightarrow l^+ l^-$ for $l = e,\, \mu$ to determine the vector-meson decay constant $f_V$. This reaction is completely described by the $f_V$ term in \eqref{eq:Lvec} with its partial decay width given by \cite{Lutz:2008km}
\begin{align}
 \Gamma_{\omega \rightarrow \,l^+ l^-} = \frac{e^4 \, f_V^2}{108 \, \pi \, m_\omega^3} \,\sqrt{1 - \frac{4 \, m_l^2}{m_\omega^2}} \left(2 m_l^2 + m_\omega^2 \right).
\end{align}
By fitting to the experimental widths for decays into both dimuon and dielectron \cite{PDG}, the decay constant is fixed to 
\begin{align}
 f_V^\omega= 140 \, \te{MeV}\,.
\end{align}
The deviation between the two values for $f_V$, $f_V^\rho$ and $f_V^\omega$, is of the order of $10\%$. This defines the accuracy which we expect to achieve.

It is interesting to compare our results to standard VMD \cite{Sakurai:1969,Landsberg:1986fd}. In this respect, a discussion based on the tree-level results \eqref{eq:kpipi} and \eqref{eq:Ftree pi} is most illuminating. As already pointed out in \cite{Ecker:1989yg}, complete VMD with the KSFR relations \cite{Kawarabayashi:1966kd,Riazuddin:1966sw} is recovered for $f_V = \sqrt{2} \, f_\pi \approx 130\,$MeV, $h_P = 2$. This would yield the common expression $m^2_{\rho,\rm bare}/(m^2_{\rho,\rm bare}-s)$ for the tree-level pion form factor in \eqref{eq:Ftree pi} and for the expression in the last brace in \eqref{eq:kpipi}. With this choice for the parameters $f_V$ and $h_P$ both coefficients $h_P^2 f_V^2 /(8 f_\pi^2)$ and $h_P f_V^2/(4 f_\pi^2)$ become 1. Our parameters in \eqref{param:fV} and \eqref{param:hp} do not agree with complete VMD. For instance, our result for $h_P$ deviates from the KSFR relation by $20\%$ which we regard as a significant difference. But our numerical value for the coefficient $h_P \, f_V^2/(4\, f_\pi^2)$, which shows up in the tree-level pion form factor \eqref{eq:Ftree pi}, is given by $\approx 1.09$. Thus our result for the pion form factor appears to be close to VMD, though our formalism with its two independent parameters $f_V$ and $h_P$ in principle allows for arbitrary deviations from VMD. As we will see in the following, there are cases where our formalism clearly deviates from VMD, even qualitatively, and there are cases where the interplay between the vector-meson parameters is such that VMD emerges. 

\section{Decays $\omega \to \pi^0 \gamma^{(\ast)}$ and $\omega \rightarrow \pi^+ \pi^- \pi^0$} \label{sec:omega decays}
In the sector of odd intrinsic parity \eqref{eq:LA} there is one parameter which needs to be determined, namely $h_A$. It will be determined in two ways: First, from the partial width of the reaction $\omega \to \pi^0 \gamma$ and, second, from the partial width of the reaction $\omega \to 3 \pi$. Resembling the case of $f_V$ we will find values which agree on the level of $10\%$. \\
Note that strictly speaking, one determines in both decay channels the combination $h_A \, f_V / \fpi$. For both determinations of $h_A$  we have used $f_V = f_V^\rho = 150 \, \te{MeV}$ extracted from the pion form factor, i.e.\ from the properties of the $\rho$-meson. Using this value is a consistent treatment because it is the $\rho$- and not the $\omega$-meson which couples to the photon in the process $\omega \to \pi^0 \gamma$ and to the two pions in the process $\omega \to 3\pi$. 

By using the $h_A$ term in \eqref{eq:LA} for the $\omega$-$\rho$-$\pi$ vertex and the $f_V$ term in \eqref{eq:Lvec} for the $\rho$-$\gamma$ vertex, the partial decay width for the decay $\omega \to \pi^0 \gamma$ is given by \cite{Terschlusen:2012xw, Terschluesen:2010ik, Lutz:2008km}
\begin{align}
 \Gamma_{\omega \rightarrow \pi^0 \gamma} = \frac{\left(m_\omega^2 - m_{\pi^0}^2 \right)^3 \, e^2}{384 \pi \, m_\omega \, m_\rho^4} \left(\frac{h_A \,f_V}{\fpi} \right)^2 . \label{eq:Gamma omega pi gamma}
\end{align}
To be in accordance with previous work \cite{Terschlusen:2012xw, Leupold:2008bp, Terschluesen:2010ik, Lutz:2008km}, $h_A$ is chosen as positive,
\begin{align}
 h_A = 2.17 \,. \label{eq:hA-pi-gamma}
\end{align}
Again, this sign is pure convention defining the pion fields $\Phi$ to produce pions. After fixing the sign of $h_A$ there is no freedom left for sign choices in other interaction terms with an odd number of pion fields. In particular, the relative sign between $n$ in the WZW structure \eqref{eq:LWZW} and $h_A$ is not a matter of convention. We will come back to this issue in section \ref{sec:pion}.

From \eqref{eq:Lvec} and \eqref{eq:LA}, one can also calculate the electromagnetic $\omega$-$\pi^0$ transition form factor. Normalized to the photon point it is given by \cite{Terschlusen:2012xw, Terschluesen:2010ik}
\begin{align}
 F_{\omega \pi^0}(q^2) &= - \, \frac{m_\rho^2}{m_\omega^2} \, \left(m_\omega^2+ q^2 \right) S_\rho(q^2) \label{eq:Ff omega pi}
\end{align}
in significant deviation from the VMD prediction \cite{Landsberg:1986fd},
\begin{align}
 F_{\omega\pi^0}^{\rm VMD}(q^2) = - m_\rho^2 \, S_\rho(q^2)\,. \label{eq:VMD-Ff omega pi}
\end{align}
Hereby, $S_\rho$ denotes the $\rho$-meson propagator,
\begin{align}
 S_\rho(q^2) &= \frac{1}{q^2 - m_\rho^2 + \, i\sqrt{q^2} \, \Gamma_\rho(q^2)} \,, \nn \\
 \Gamma_\rho(q^2) &= \left( \frac{q^2 - 4m_{\pi^+}^2}{m_\rho^2 - 4  m_{\pi^+}^2} \right)^{\hspace{-0.2em} 3/2} \frac{m_\rho^2}{q^2} \, \Gamma_{\rho,0} \label{eq:rho-width full}
\end{align}
with the on-shell $\rho$-meson width $\Gamma_{\rho,0} = 150 \, \te{MeV}$ \cite{PDG}. As discussed in \cite{Terschlusen:2012xw, Terschluesen:2010ik}, equation \eqref{eq:Ff omega pi} provides a much better description of the available data than VMD\footnote{In \cite{Terschlusen:2012xw, Terschluesen:2010ik}, we have included an additional flavor-breaking interaction term (see discussion in section \ref{sec:fudi}). It is numerically much less important. Although we include in the present work only the $h_A$ term as the numerically most important contribution to the transition $\omega \rightarrow \pi^0$, the value for $h_A$ differs less than $10\%$ from the previous determinations in \cite{Terschlusen:2012xw, Terschluesen:2010ik, Lutz:2008km}.}. Note that the $\omega$-$\pi^0$ transition form factor has also been considered in \cite{Schneider:2012ez}.

Obviously, our vector-meson coupling constants $f_V$ and $h_A$ enter the result \eqref{eq:Gamma omega pi gamma} for the partial decay width of the $\omega$-meson, whereas the $\omega$ transition form factor \eqref{eq:Ff omega pi} is independent of the coupling constants. At tree level, i.e.\ ignoring the $\rho$-meson width in the propagator, the result is even independent of any parameter that encodes interactions with vector mesons. This qualitative feature is shared with VMD, cf.\ \eqref{eq:VMD-Ff omega pi}, but our result is qualitatively and quantitatively different from VMD. This is in contrast to the previously discussed case of the pion form factor where a result numerically very close to VMD has been obtained. In the latter case the result emerged as an interplay between two contributions: There is one direct coupling of pions to a photon via the electric charge of the pion. This contribution emerges from the $\chi$PT Lagrangian \eqref{eq:LChPT}. In addition, there is a second contribution from the vector-meson Lagrangian \eqref{eq:Lvec} leading to the term $\sim h_P  f_V^2$ in \eqref{eq:Ftree pi}. The interplay of these two terms and the particular numerical values for the vector-meson coupling constants lead to a result that is numerically close to VMD. For the $\omega$ transition form factor such an interplay cannot appear. For this process there is only one contribution in our formalism, not a combination of a direct term and a vector-meson term. In spite of the fact that this single contribution is a two-step process with a virtual $\rho$-meson in the intermediate state, the result is not of VMD type. The reason is the particular vertex and propagator structure in the utilized tensor-field representation. We will come back to the VMD discussion in section \ref{sec:pion} where we will discuss the pion transition form factor.

The decay $\omega \rightarrow 3\pi$ was already calculated in \cite{Leupold:2008bp} by one of the authors. In this publication, we will additionally consider the influence of pion-pion rescattering and of lifting the isospin limit \cite{Strandberg:2012}. Furthermore, in \cite{Leupold:2008bp} the full vector-meson Lagrangian from \cite{Lutz:2008km} was used whereas a simplified version is used here (see discussion in section \ref{sec:fudi}). Recently, this decay was brought into focus again by theoretical calculations using dispersion theory \cite{Niecknig:2012sj} and ongoing experimental Dalitz-plot investigations \cite{Heijkenskjoeld:WASA, Heijkenskjoeld:KLOE}.

Following the notation in \cite{Leupold:2008bp}, the double-differential decay width 
\begin{align}
 \frac{\dint^2 \Gamma_{\omega \rightarrow 3\pi}}{\dint m_{12}^2 \, \dint m_{23}^2} = \frac{1}{(2 \pi)^3}  \, \frac{P_{3 \pi}}{32\, m_\omega^3} \left| C_{\omega \rightarrow 3\pi} \right|^2 \label{eq:omega3pi width}
\end{align}
includes the reduced matrix element $C_{\omega \rightarrow 3\pi}$ and the p-wave phase-space factor
\begin{align}
  P_{3 \pi} = - \frac{1}{3} \, \eps_{\mu\nu\alpha\beta} \, p^\mu p_1^\nu p_2^\alpha \, \eps_{\bar{\mu}\bar{\nu}\bar{\alpha}}^{\phantom{\bar{\mu}\bar{\nu}\bar{\alpha}}\, \beta} \, p^{\bar{\mu}} p_1^{\bar{\nu}} p_2^{\bar{\alpha}} \,. \label{eq:P3pi}
\end{align}
Thereby, $p$, $p_1$, $p_2$ and $p_3$ denote the four-momenta of the incoming $\omega$-meson and the outgoing $\pi^+$-, $\pi^-$- and $\pi^0$-mesons, respectively. The pion momenta are collected in the three invariant masses $m_{ij}^2 := (p_i +p_j)^2$ fulfilling
\begin{align}
 m_{12}^2 + m_{23}^2 + m_{13}^2 = 2 m_{\pi^+}^2 + m_{\pi^0}^2 + m_\omega^2 \,. \label{eq:def m13}
\end{align}
Besides other aspects we will study in the following the two cases where the physical (isospin breaking) pion masses are used in \eqref{eq:omega3pi width} - \eqref{eq:def m13} and where an isospin averaged pion mass is used.

\begin{figure}[t]
 \centering
 \includegraphics[width = 0.15\textwidth]{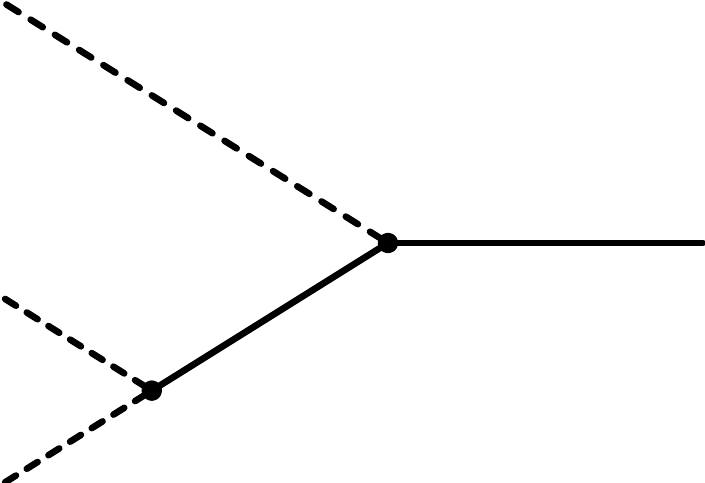} \hspace{2em}
 \includegraphics[width = 0.214283\textwidth]{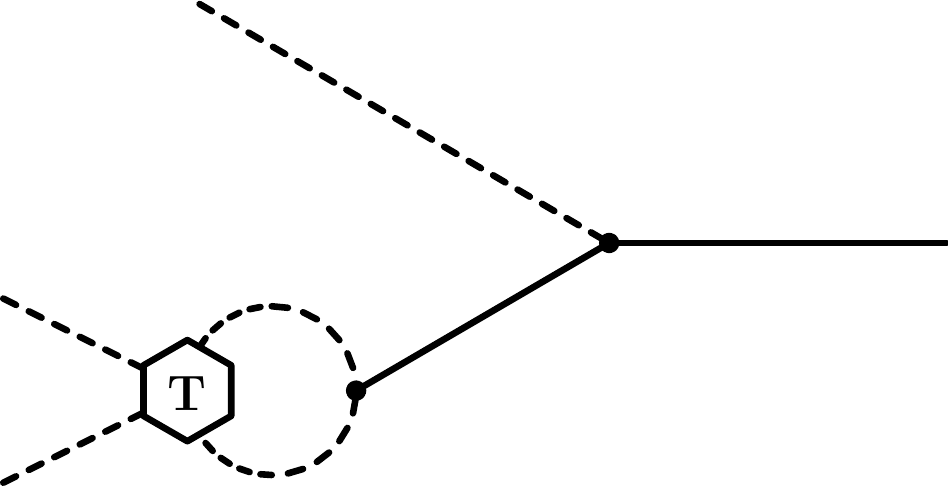} 
 \caption{Relevant process for the decay $\omega \rightarrow 3\pi$ without (\textbf{left}) and with rescattering (\textbf{right}). The dashed lines represent the pions, the solid ones the vector mesons. See also Fig. \ref{fig:scatt-kernel}.}
 \label{fig:omega3pi}
\end{figure}

The decay $\omega \rightarrow 3\pi$ can happen via a virtual $\rho^0$-, $\rho^+$- or $\rho^-$-meson whereby the $\omega$-$\rho$-$\pi$ vertex is described by the $h_A$ term in \eqref{eq:LA} and the $\rho$-$2\pi$ vertex by the $h_P$ term in \eqref{eq:Lvec} yielding the reduced matrix element \cite{Leupold:2008bp}
\begin{align}
 C_{\omega \rightarrow 3 \pi} = \frac{f_V \, h_A \, h_P}{4 \, \fpi^3 \, m_\omega} \, \sum_{\substack{(i,j) = (1,2),\\ (2,3), (1,3)}} \hspace{-1em} S_\rho(m_{ij}^2) \left(m_{ij}^2 + m_\omega^2 \right). \label{eq:Matrixelemen omega3pi}
\end{align}
In the following, we evaluate \eqref{eq:omega3pi width} in three different ways: First, we approximate the $\rho$-meson propagator in \eqref{eq:Matrixelemen omega3pi} by the tree-level expression, i.e. without any $\rho$-meson width,
\begin{align}
 \left[S_\rho(q^2)\right]^{-1} \approx q^2 - m_\rho^2\,. \label{eq:rho-width approx} 
\end{align}
This approach will be labeled ``t.l.''. Second, we use the form \eqref{eq:rho-width full} where a Breit-Wigner width was added manually (labeled by ``B.W.''). Third, we will include rescattering of the two pions in the $\rho$-meson channel in the same way as done for the pion form factor in the previous section (labeled with ``resc.''). For the latter, one has to add up the tree-level contribution (left-hand side in Fig. \ref{fig:omega3pi}) and the contribution from rescattering (right-hand side in Fig. \ref{fig:omega3pi}) by replacing the $\rho$-meson propagator in \eqref{eq:Matrixelemen omega3pi} according to
\begin{align}
 S_\rho(s)& \nn \\
 \longmapsto \ &\frac{1}{s - m_{\rho,\rm bare}^2} \left\{ 1 + \left( J(s) - \te{Re}\,J(\mu^2) \right) t_{l=1, I=1}(s) \right\} \nn \\
 &= \, \frac{1}{s - m_{\rho,\rm bare}^2} \ \frac{1}{1 - k_{l=1, I=1}(s) \left( J(s) - \te{Re}\,J(\mu^2) \right)}  \label{eq:resc-prop}
\end{align}
with the renormalization point $\mu = m_\rho = m_{\rho,\te{bare}}$ (cf. discussion after \eqref{eq:def J}).

The approach including rescattering is used to fix the open parameter $h_A$ as
\begin{align}
 h_A = 2.02 \,. \label{eq:hA-3pi}
\end{align}
This result has been obtained in the isospin limit, i.e. for an averaged $\pi^{\pm}$/$\pi^0$ mass. However, up to our accuracy the result for $h_A$ will be the same if distinct pion masses are used instead in the phase-space calculation. The result \eqref{eq:hA-3pi} obtained from $\omega \rightarrow 3 \pi$ should be compared with \eqref{eq:hA-pi-gamma} as obtained from $\omega \rightarrow \pi^0 \gamma$. Again, we observe satisfying agreement on the level of $10\%$.

In a next step, we explore the differences between the three approaches. Note that we have determined $h_A$ from the rescattering approach because we regard it as the most trustworthy scheme. There, the width and, in accordance with analyticity, a corresponding real part is generated and not added by hand. On the other hand, while the inclusion of a $\rho$-meson width is reasonable from a physics point of view, it is technically not mandatory since the invariant mass of the two-pion states is always below the $\rho$-meson mass for the decay $\omega \rightarrow 3\pi$. Therefore, it is worth to explore the quantitative differences between the three approaches. 

In Tab. \ref{tab:omega results}, the results for the partial decay width calculated with the tree-level approximation of the $\rho$-meson propagator \eqref{eq:rho-width approx} and with a manually added Breit-Wigner width \eqref{eq:rho-width full}, respectively, are listed for both averaged and distinct pion masses. The calculations with the tree-level approximation differ only by about $5.5\%$ from the result with rescattering whereas the calculations with the Breit-Wigner width differ by about $7.5\%$. Thus, for the decay $\omega \rightarrow 3\pi$ it is obviously not enough to simply include a Breit-Wigner width in the $\rho$-meson propagator to account for rescattering in the final channels. On the other hand, the influence of distinct pion masses is less than $1\%$ in the calculation without width and about $2\%$ in the one with width. 
\begin{table}[h]
 \caption{Partial decay width for the decay $\omega \rightarrow 3\pi$ with the tree-level approximation of the $\rho$-meson propagator and with the manually added Breit-Wigner width compared to the experimental value \cite{PDG} and calculated both for an averaged pion mass and distinct pion masses.}
 \label{tab:omega results}
 \begin{tabular}{l|c|c}
 & averaged mass & distinct masses \\ \hline
 $\Gamma_{\omega \rightarrow 3\pi}^{\hspace{0.1em} \text{t.l.}}$ [MeV] & 7.21 & 7.15 \\[0.3em]
 $\Gamma_{\omega \rightarrow 3\pi}^{\hspace{0.1em} \text{B.W.}}$ [MeV] & 7.07 & 7.01 \\[0.3em]
 $\Gamma_{\omega \rightarrow 3\pi}^{\rm \hspace{0.1em} exp. / resc.}$ [MeV] & \multicolumn{2}{c}{$7.57 \pm 0.01$} \end{tabular}
\end{table}

\begin{figure}[h]
 \centering
 \includegraphics[width = 0.3\textwidth, trim = 50 0 50 0]{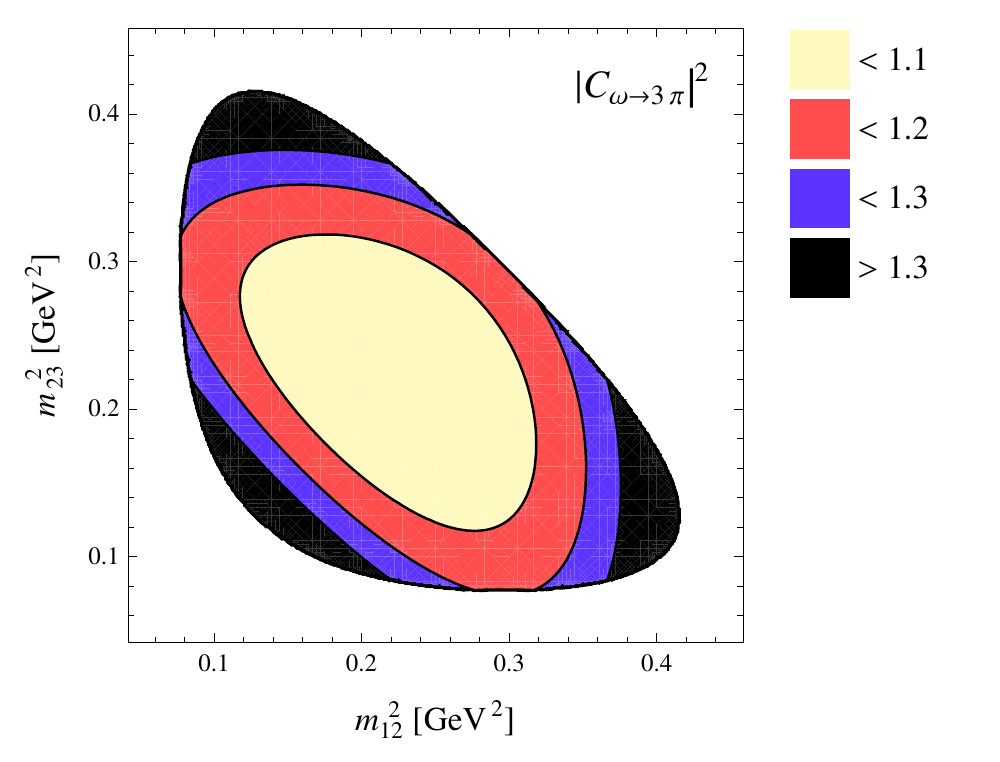}
\caption{Dalitz plot of the reduced matrix element \eqref{eq:Matrixelemen omega3pi} for the decay $\omega \rightarrow 3 \pi$ calculated with pion-pion rescattering, averaged pion masses and normalized to the Dalitz-plot center.}
 \label{fig:DalitzPlots}
\end{figure}

Having determined the last free parameter from integrated decay data, we turn now to the differential decay width of $\omega \rightarrow 3 \pi$. In Fig. \ref{fig:DalitzPlots}, the Dalitz plot of the reduced matrix element \eqref{eq:Matrixelemen omega3pi} is plotted for an averaged pion mass. As already stated in \cite{Leupold:2008bp}, the squared reduced matrix element $|C_{\omega \rightarrow 3\pi}|^2$ is rather flat in the kinematically allowed region. For the plot, the matrix element was calculated including rescattering. However, neither the calculations with the tree-level approximation of the $\rho$-meson propagator and with a manually added Breit-Wigner width nor the calculation with distinct pion masses differ much from the plot shown.

To quantify the Dalitz plot, we follow \cite{Niecknig:2012sj} and introduce variables
\begin{align}
 &x := \frac{m_{23}^2 - m_{13}^2}{\sqrt{3} \, R_\omega}\, = \sqrt{z} \, \cos\varphi \,, \nn \\
 &y := \frac{s_0 - m_{12}^2}{R_\omega}\, = \sqrt{z} \, \sin\varphi \,, \nn \\
 &R_\omega := \frac{2}{3} \, m_\omega \left(m_\omega - 3 m_\pi\right), \ s_0 = \frac{1}{3} \left(m_\omega^2 + 3 m_\pi^2 \right)
\end{align}
with the polar coordinates $z$ and $\varphi$. Herewith, the squared reduced matrix element can be expressed as
\begin{align}
 \left| C_{\omega \rightarrow 3\pi}(z,\varphi) \right|^2 = & \left|\mathcal{N}\right|^2 \left\{ 1 + 2 \alpha z + 2 \beta z^{3/2} \, \sin 3\varphi + 2 \gamma \, z^2 \right. \nn \\
 &\left. {} + 2 \delta z^{5/2} \, \sin 3 \varphi + \mathcal{O}(z^3) \right\} \label{eq:param}
\end{align}
including the normalization $\mathcal{N}$ to the Dalitz-plot center $x = y =0$. In the following, we terminate the $z$-$\varphi$ expansion at the Dalitz-plot parameters $\alpha$, $\beta$, $\gamma$ or $\delta$ and call the obtained parametrization $|C_{\rm pol}|^2$. Following the approach in \cite{Niecknig:2012sj}, the Dalitz-plot parameters are obtained by minimizing
\begin{align}
 \chi^2 =\,& \frac{1}{\mathcal{N}_{\mathcal{D}}} \int_\mathcal{D} \dint z \,\dint \varphi \, \left(P_{3\pi}(z,\varphi) / P_{3\pi}(0,0) \right)^2 \nn \\
 &\phantom{\frac{1}{\mathcal{N}_{\mathcal{D}}} \int_\mathcal{D}} \cdot \left(  \frac{ \left| C_{\rm pol}(z,\varphi) \right|^2 - \left|C_{\omega \rightarrow 3 \pi}(z,\varphi) \right|^2}{\left| \mathcal{N} \right|^2} \right)^2 , \nn \\
 \mathcal{N}_\mathcal{D} =\,& \int_{\mathcal{D}} \dint z\, \dint \varphi \label{eq:chi2}
\end{align}
while integrating over the kinematically allowed region of the Dalitz plot, $\mathcal{D}$. Thereby, $\sqrt{\chi^2}$ denotes the average deviation of the polynomial description from the matrix element $P_{3\pi} \left| C_{\omega \rightarrow 3 \pi} \right|^2$ relative to the Dalitz plot center. As the outer parts of the Dalitz plot are expected to be statistically less important in an experiment, the full matrix element is used instead of the reduced one for the minimization procedure to give less weight to these parts. The resulting parameters for an averaged pion mass are collected in Tab. \ref{tab:Dalitz-plot parameters}. We see, first of all, that the parameter $\alpha$ is significantly larger than the other parameters. Its value is more or less the same for the three studied approaches. Physically the sign of $\alpha$, i.e.\ a rise of the matrix element towards the boundaries of the Dalitz plot (see also Fig. 4), is related to the intermediate $\rho$-meson: For larger invariant masses of a pion pair the $\rho$-meson propagator becomes larger, because one approaches the $\rho$-meson peak. Of course, the peak itself lies outside of the kinematically accessible region. For the other three parameters $\beta$, $\gamma$ and $\delta$ included in the fits one observes significant deviations between the three approaches. We stress again that we regard the rescattering scenario as the most reliable one since it respects unitarity and analyticity for the incorporated two-pion rescattering \cite{Leupold:2009}. 
\begin{table}[h]
 \caption{Dalitz-plot parameters and $\sqrt{\chi^2}$ \eqref{eq:chi2} performed with the reduced matrix element \eqref{eq:Matrixelemen omega3pi} including the tree-level approximation of the $\rho$-meson propagator, a manually inserted Breit-Wigner width and pion-pion rescattering, respectively, performed with an averaged pion mass.}
 \label{tab:Dalitz-plot parameters}
 \begin{tabular}{l|c|c|c|c|c}
  & $\alpha \cdot 10^3$ & $\beta \cdot 10^3$ & $\gamma \cdot 10^3$ & $\delta \cdot 10^3$ & $\sqrt{\chi^2} \cdot 10^3$ \\ \hline
 t.l.& $226$ & -- & -- & -- & $9.6$ \\
 B.W. & $193$ & -- & -- & -- & $6.0$ \\
 resc.& $202$ & -- & -- & -- & $6.6$ \\ \hline
 t.l. & $209$ & $77$ & -- & -- & $3.5$ \\
 B.W. & $182$ & $49$ & -- & -- & $1.9$ \\
 resc. & $190$ & $54$ & -- & -- & $2.1$ \\ \hline
 t.l. & $180$ & $60$ & $83$ & -- & $0.9$ \\
 B.W. & $166$ & $40$ & $46$ & -- & $0.3$ \\
 resc. & $172$ & $43$ & $50$ & -- & $0.4$ \\ \hline 
 t.l. & $185$ & $40$ & $66$ & $48$ & $0.2$ \\
 B.W. & $168$ & $33$ & $40$ & $16$ & $0.0$ \\
 resc. & $174$ & $35$ & $43$ & $20$ & $0.1$
 \end{tabular}
\end{table}

It is reasonable to compare our approach and our results for the Dalitz-plot parameters to \cite{Niecknig:2012sj}. Our approach is based on the Lagrangian proposed in section \ref{sec:ind}. Parameters have been fitted to various observables and cross-checked against each other. The tree-level results are improved by a Bethe-Salpeter equation for the rescattering of the pions that emerge from a $\rho$-meson (see Fig.\ \ref{fig:omega3pi}). What is not included in our approach is cross-channel rescattering, i.e.\ the scattering of the third pion with one of the two that emerged from the $\rho$-meson. The approach of \cite{Niecknig:2012sj} is based on a dispersive treatment of the formal reaction amplitude $\omega + \pi \to 2 \pi$. It is assumed that the imaginary part of this amplitude is governed by the two-pion intermediate state and that the effect of other channels like $K \, \bar K$ or the elastic one, $\omega \, \pi,$ is negligible. Assuming an appropriate high-energy behavior of the such approximated amplitude one obtains the full amplitude by a dispersion relation. There is one undetermined parameter, the initial strength of the $\omega-3\pi$ vertex. This parameter is fixed from the integrated decay width. Rescattering and cross-channel rescattering of all three pions can be included in this dispersive framework. The necessary input is the pion-pion scattering amplitude which is not obtained from a microscopic model, as in our case, but is taken from recent data analyses. Comparing the Dalitz-plot parameters we note first of all that we observe qualitative agreement concerning the signs of all parameters and concerning the fact that $\alpha$ is the dominant one. Overall our parameters are larger than the ones obtained in \cite{Niecknig:2012sj}. 
Clearly there is one aspect where our approach is less powerful: We do not account for cross-channel rescattering. On the other hand, we have predictive power concerning absolute numbers (determining the parameter $h_P$ from $\rho \rightarrow 2 \pi$ and the parameter $h_A$ in principle from $\omega \rightarrow \pi^0 \gamma$). This becomes even more obvious in the scattering process $e^+ e^- \rightarrow 3 \pi$ where we can predict the size of the cross section (see section \ref{sec:scattering}). As already spelled out in \cite{Niecknig:2012sj} the dispersive framework cannot predict this cross section, but could predict a Dalitz distribution for a given invariant mass $\sqrt{s} = m_{e^+ e^-}$. 

%
%
\section{Pion transition form factor} \label{sec:pion}

\begin{figure}[h]
\begin{centering}
 \includegraphics[trim = 0 0 -80 0, width = 0.2\textwidth]{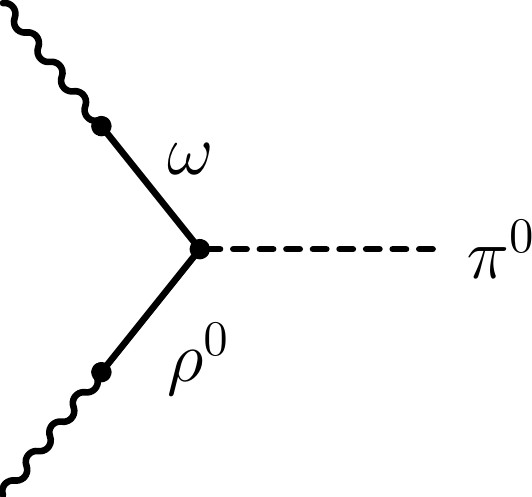}
 \includegraphics[trim = -80 0 0 0, width = 0.2\textwidth]{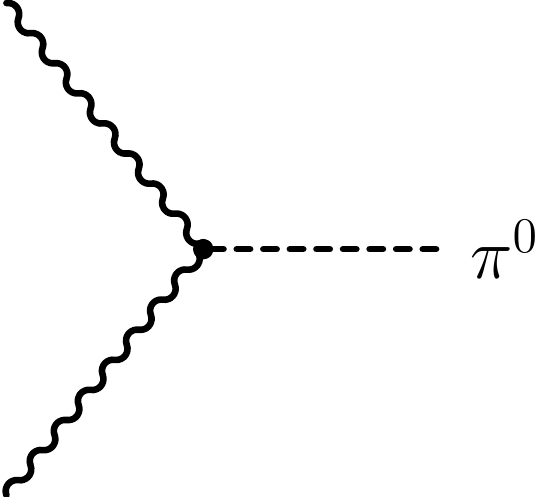}
 \caption{\textbf{Left:} Indirect decay $\pi^0 \rightarrow \gamma^{(\ast)} \gamma^{(\ast)}$ via a virtual $\rho$- and a virtual $\omega$-meson. \textbf{Right:} Direct decay via the WZW Lagrangian.}
 \label{fig:pion-decay-feyn}
\end{centering}
\end{figure}

In our approach the decay of a pion into two (real or virtual) photons can happen in two ways, either via a virtual $\rho$- and a virtual $\omega$-meson or with a direct $\pi$-$2\gamma$ coupling (see left- and right-hand side in Fig. \ref{fig:pion-decay-feyn}, respectively). For the indirect decay via virtual vector mesons, the necessary parameters $h_A$ of the $2V$-$\pi$ vertex and $f_V$ of the $V$-$\gamma^{(\ast)}$ vertex have already been determined in the previous sections. The direct contribution is described by the WZW Lagrangian \eqref{eq:LWZW}. The modulus of the parameter $n$ is given by the number of colors, $|n| = N_c = 3$ \cite{Witten:1983tw}. There is no sign choice for $n$ left since the freedom in the convention of the pion fields was already used to fix the sign of $h_A$ as positive. We will determine the sign of $n$ in this section by comparing the single-virtual pion transition form factor to space-like data and use the result to predict the partial decay widths for the decays into one and two dielectrons. Thereby, the decay of a virtual photon into a dielectron is described by usual quantum electrodynamics. 
%

%
%
%
%
The general matrix element for the single-virtual decay, i.e. the decay into a dielectron and a real photon, can be expressed as \cite{Landsberg:1986fd}
\begin{align}
 \mathcal{M}_{\pi^0 \rightarrow \gamma \, e^+ e^-} =& - e^2 f_{\rm s.v.}(q^2) \, \eps^{\mu\nu\alpha\beta} \, q_\mu k_\nu \, \eps^\ast_\alpha(k, \lambda_\gamma) \nn \\
 &{} \cdot \frac{1}{q^2} \, \bar{u}(q_1, \lambda_1) \, \gamma_\beta \, v(q_2, \lambda_2)
\end{align}
including the single-virtual pion transition form factor $f_{\rm s.v.}(q^2)$. Here, $q$ and $k$ are the four-momenta of the outgoing real and virtual photon, respectively. $u(q_{1}, \lambda_{1})$ and $v(q_{2}, \lambda_{2})$ denote the wave functions of the electron and the positron with their respective four-momenta $q_{1}$ and $q_{2}$ and $\eps^\ast_\alpha(k, \lambda_\gamma)$ is the wave function of the outgoing (real) photon. 

The form factor is normalized to its value at the photon point $q^2 = 0$,
\begin{align}
 F_{\rm s.v.}(q^2) := \frac{f_{\rm s.v.}(q^2)}{f_{\rm s.v.}(0)} \,, \label{eq:Ff-sv-gen}
\end{align}
such that $F_{\rm s.v.}(0) =1$.

The pion transition form factor includes both the vector-meson contribution calculated with the Lagrang\-ians \eqref{eq:Lvec} and \eqref{eq:LA} and the direct contribution calculated with the WZW Lagrangian \eqref{eq:LWZW},
\begin{align}
 &f_{\rm s.v.} = f_{\rm s.v.}^{\rm vec} + f_{\rm s.v.}^{\rm WZW} \, , \nn \\ 
 &f_{\rm s.v.}^{\rm vec}(q^2) = \frac{e \, f_V^2 \, h_A}{6 \,\fpi} \left( \frac{1}{m_\rho^2}\,S_\omega(q^2) + \frac{1}{m_\omega^2} \, S_\rho(q^2) \right) q^2 \,, \nn \\
 &f_{\rm s.v.}^{\rm WZW}(q^2) = \frac{n \, e}{ 12 \, \pi^2 \fpi}\,. \label{eq:Ff-sv}
\end{align}
For the space-like region ($q^2 < 0$) and for the decays $\pi^0 \rightarrow \gamma^{(*)} \gamma^{(*)}$ the phase space is closed for all relevant decays which provide a width for the vector mesons. Therefore, we approximate the vector-meson propagators by their tree-level expression,
\begin{align}
 \left[S_V(q^2)\right]^{-1} \approx q^2 - m_V^2\,. 
\end{align}

For the decay into two real photons,
\begin{align}
 \Gamma_{\pi^0 \rightarrow 2 \gamma} = \frac{m_{\pi^0}^3}{64 \pi} \, \left|e f_{\rm s.v.}(0) \right|^2 \,, \label{eq:G2gamma}
\end{align}
the form factor is only needed at the photon point. Since $f_{\rm s.v.}^{\rm vec}(0) = 0$, the full form factor at the photon point is equal to the WZW contribution. Hence, the decay width is proportional to $n^2$ and cannot be used to determine the sign of the parameter $n$ in the WZW Lagrangian \eqref{eq:LWZW}. It is in very good agreement with the experimental data \cite{PDG}, 
\begin{align}
 \Gamma_{\pi^0 \rightarrow 2 \gamma} &= 7.79 \cdot 10^{-9} \,\te{GeV}, \label{res:G2g} \\ 
 \Gamma_{\pi^0 \rightarrow 2 \gamma}^{\rm \, exp.} &= (7.63 \pm 0.16) \cdot 10^{-9} \,\te{GeV}. 
\end{align}

\begin{figure}[h]
 \begin{centering}
 \includegraphics[trim = 40 50 0 60, width = 0.5\textwidth]{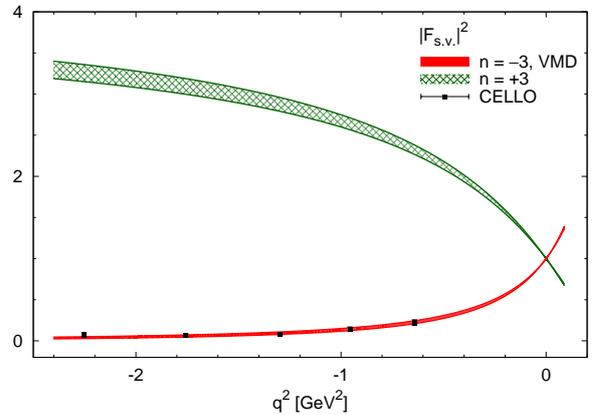}  
  \caption{Squared normalized $\pi$-$\gamma$ transition form factor compared to the VMD prediction and to space-like data taken by the CELLO collaboration \cite{Behrend:1990sr} for $n = - 3$ (solid red band) and $n = +3$ (dashed green band).}
  \label{fig:Ff-pi-space-like}
 \end{centering}
\end{figure}

\begin{figure}[h]
 \begin{centering}
 \includegraphics[trim = 40 50 0 60, width = 0.5\textwidth]{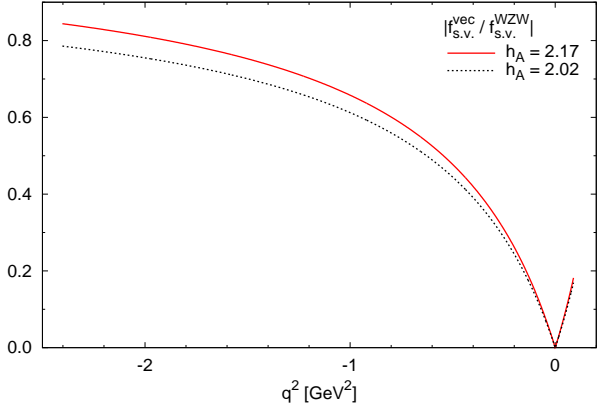}
  \caption{Modulus of the ratio $f_{\rm s.v.}^{\rm vec}(q^2) / f_{\rm s.v.}^{\rm WZW}(q^2)$.}
  \label{fig:Ff-pi-ratio}
 \end{centering}
\end{figure}

To determine the sign of the WZW parameter $n$, the normalized form factor 
\begin{align}
 &F_{\rm s.v.}(q^2) \nn\\
& = 1 - \,\frac{2 \pi^2 \, f_V^2 \, h_A}{n} \left( \frac{q^2}{m_\omega^2\left(m_\rho^2 - q^2 \right)} \, +\, \frac{q^2}{m_\rho^2 \left( m_\omega^2 -q^2 \right)} \right) \label{eq:Ff-sv-approx}
\end{align}
is compared to experimental data in the space-like region taken by the CELLO collaboration \cite{Behrend:1990sr} in Fig. \ref{fig:Ff-pi-space-like}. The error bands for $n = -3$ (solid red band) and $n = +3$ (dashed green band) correspond to the variation of $h_A$ between $2.02$ and $2.17$ for fixed $f_V = f_V^\rho = 150 \, \te{MeV}$. We recall that these values for $h_A$ have been determined in \eqref{eq:hA-pi-gamma} and \eqref{eq:hA-3pi} from $\omega \rightarrow \pi^0 \gamma$ and $\omega \rightarrow 3\pi$, respectively. Obviously, a negative sign is needed to describe the data\footnote{We get the opposite sign to \cite{Scherer:2002tk} due to different conventions of the pion fields. What matters, of course, is the relative sign between $n$ and $h_A$. We could have used a positive $n$ and a negative $h_A$. But we want to stay in agreement with the convention of \cite{Terschlusen:2012xw, Leupold:2008bp, Terschluesen:2010ik, Lutz:2008km}.}, 
\begin{align}
 n &= - N_c = - 3\,.
\end{align}
To evaluate the relative importance of vector and WZW contribution in the form factor, the modulus of the ratio $f_{\rm s.v.}^{\rm vec}(q^2) / f_{\rm s.v.}^{\rm WZW}(q^2)$ is plotted in Fig. \ref{fig:Ff-pi-ratio}. In the decay region ($0 \leq q^2 \leq m_{\pi^0}^2$), the vector contribution is less important than the direct WZW contribution, it is at most about $3\%$ of the WZW contribution. Nevertheless, the importance of the vector contribution increases with growing $|q^2|$ both in the time- and space-like region.

\begin{figure}[h]
 \begin{centering}
 \includegraphics[trim = 40 50 0 60, width = 0.5\textwidth]{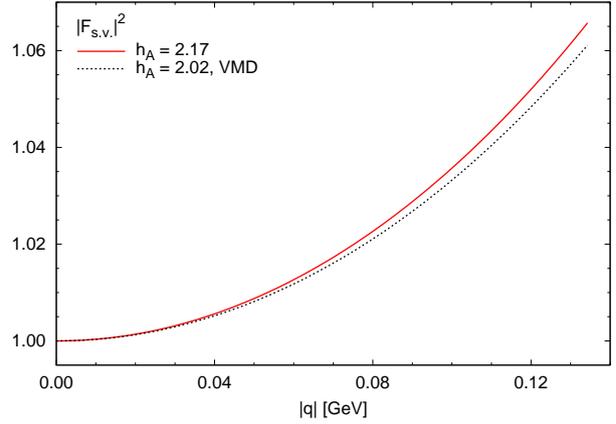} 
  \caption{Squared normalized $\pi$-$\gamma$ transition form factor in the kinematically allowed time-like decay region for $n = -3$ and for $h_A = 2.17$ (red solid line) and $h_A = 2.02$ (dashed black line), respectively, compared to the VMD prediction.}
  \label{fig:Ff-pi-decay}
 \end{centering}
\end{figure}

In contrast to \eqref{eq:Ff-sv-approx}, the form factor predicted by VMD is given by
\begin{align}
 F^{\rm VMD}_{\rm s.v.}(q^2) = \frac{1}{2} \left( \frac{m_\rho^2}{m_\rho^2 - q^2} + \frac{m_\omega^2}{m_\omega^2 - q^2} \right). \label{eq:Ff-pi-VMD}
\end{align}
The VMD prediction lies within the error band for $n = -3$ in Fig. \ref{fig:Ff-pi-space-like}. Additionally, both the form factor \eqref{eq:Ff-sv-approx} for $h_A = 2.17$ and $h_A = 2.02$ and the VMD prediction \eqref{eq:Ff-pi-VMD} are plotted in the kinematically allowed time-like decay region $2 m_e \leq \sqrt{q^2} \leq m_{\pi^0}$ in Fig. \ref{fig:Ff-pi-decay}. There is only a small deviation between our prediction and the VMD prediction visible and for $h_A = 2.02$ both predictions lie on top of each other. Indeed, for $m_\omega \approx m_\rho \approx m_V$ the predictions would be the same if $4 \pi^2 f_V^2 \, h_A / |n| = \,m_V^2\,$. The obtained results $(802 \,\te{MeV})^2$ and $(773 \,\te{MeV})^2$, respectively, are close to the averaged $\rho$/$\omega$ mass, $779.5 \, \te{MeV}$, and explain the numerically good agreement of the two predictions. Like for the pion form factor (cf.\ the discussion at the end of section \ref{sec:coupling constants}) we observe an interplay of two terms which conspire such that effectively a VMD type result emerges. One term comes from the respective lowest-order $\chi$PT Lagrangian, \eqref{eq:LChPT} and \ref{eq:LWZW}, respectively, and the other term from the vector mesons. We recall, however, that there are also cases where no VMD type behavior emerges, notably the $\omega$-$\pi^0$ transition form factor discussed in section \ref{sec:omega decays} and in \cite{Terschluesen:2010ik}.

By fixing the relative sign between $n$ and $h_A$ in the space-like region we have obtained full predictive power for the time-like region of the pion transition form factor. In the following we will study the decay region, i.e. \ $q^2 < m_{\pi^0}^2$ where the processes $\pi^0 \rightarrow \gamma e^+ e^-$ and $\pi^0 \rightarrow 2e^+ 2e^-$ appear. The production process $e^+ e^- \rightarrow \pi^0 \gamma^{(\ast)}$ will not be studied in the present work. 

The double-differential decay rate for the process $\pi^0 \rightarrow \gamma \,e^+ e^-$ is given by \cite{PDG}
\begin{align}
 \frac{\te{d}^2 \Gamma_{\pi^0 \to \gamma \, e^+ e^-}}{\te{d}q^2 \, \te{d}m_{e^+ \gamma}^2} = \frac{e^4}{(2 \pi)^3} \, \frac{1}{32 m_{\pi^0}^3} \, \left| f_{\rm s.v.}(q^2) \right|^2 \,\frac{P_{2l}}{q^4} \label{eq:dddr}
\end{align}
including the invariant mass $m_{e^+ \gamma}^2 := (q_2 + k)^2$ and the phase-space factor
\begin{align}
 P_{2l} =& - \eps^{\mu\nu\alpha\beta} \, k_\mu \,q_\nu \, \eps^{\bar{\mu}\bar{\nu}\bar{\alpha}}_{\phantom{\bar{\mu}\bar{\nu}\bar{\alpha}}\beta} \, k_{\bar{\mu}} \, q_{\bar{\nu}} \nn \\
 & \cdot \hspace{-0.5em} \sum_{\lambda_1,\, \lambda_2} \bar{u}(q_1, \lambda_1) \, \gamma_\alpha \, v(q_2,\lambda_2)  \, \bar{v}(q_2,\lambda_2) \, \gamma_{\bar{\alpha}} \, u(q_1,\lambda_1) \,.
\end{align}
By integrating Eq. \eqref{eq:dddr} in the kinematically allowed time-like decay region, the single-differential decay rate is obtained \cite{Landsberg:1986fd},
\begin{align}
 \frac{\te{d} \Gamma_{\pi^0 \rightarrow \gamma \, e^+ e^-}}{\te{d}q^2 \, \Gamma_{\pi^0 \rightarrow 2\gamma}} =& \, \frac{e^2}{6 \pi^2} \, \frac{1}{q^2} \, \sqrt{1 - \frac{4 m_e^2}{q^2}} \left(1 + \frac{2m_e^2}{q^2} \right) \nn \\
 & \cdot \left(1 - \frac{q^2}{m_{\pi^0}^2} \right)^3 \left| F_{\rm s.v.}(q^2) \right|^2 .  \label{eq:sddr}
\end{align}
It is normalized to the decay width for the decay into two real photons \eqref{eq:G2gamma}.

\begin{figure}[h]
 \begin{centering}
 \includegraphics[trim = 40 50 0 60, width = 0.5\textwidth]{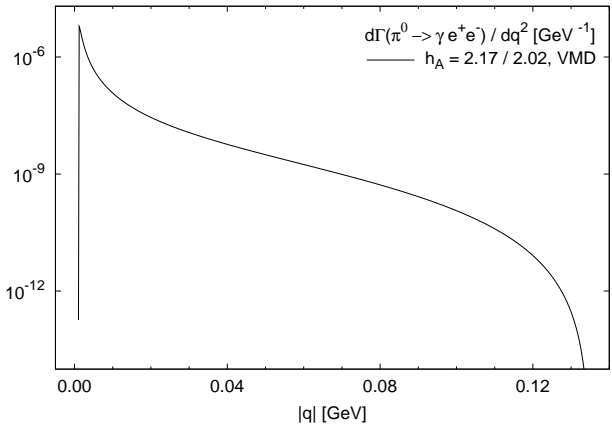}
 \caption{Single-differential decay width for the decay $\pi^0 \rightarrow \gamma \, e^+ e^-$.}
 \label{fig:pi-dielectron-sd}
 \end{centering}
\end{figure}

In Fig. \ref{fig:pi-dielectron-sd}, the single-differential decay width obtained from Eq. \eqref{eq:sddr} is shown. Here, there is no difference between our approach for the different values of $h_A$ and VMD visible. For calculating the VMD width, Eq. \eqref{eq:sddr} was used together with the experimental width for the decay into two real photons. 

Integrating the single-differential decay width yields for $h_A = 2.17$ and $f_V = 150 \, \te{MeV}$ the partial width
\begin{align}
 \Gamma_{\pi^0 \rightarrow \gamma \, e^+ e^-} = 9.26 \cd 10^{-11} \,\te{GeV} 
\end{align}
which is in very good agreement with the experimental value
\begin{align}
 \Gamma_{\pi^0 \rightarrow \gamma \, e^+ e^-}^{\rm \, exp.} = (9.07 \pm 0.33) \cd 10^{-11} \,\te{GeV} \,.
\end{align}
The result for $h_A = 2.02$ differs less than $1\%$ from the one obtained with $h_A = 2.17$. Since both the uncertainties in $h_A$ and in $f_V$ are about $10\%$, approximately the same variation in the result is achieved if one takes $h_A \,f_V / \fpi$ as determined form $\omega \to \pi^0 \gamma$ and uses $f_V$ determined from $\omega \to e^+ e^-$. 

\begin{figure}[h]
 \begin{centering}
 \includegraphics[trim = 0 0 0 -10, width = 0.15\textwidth]{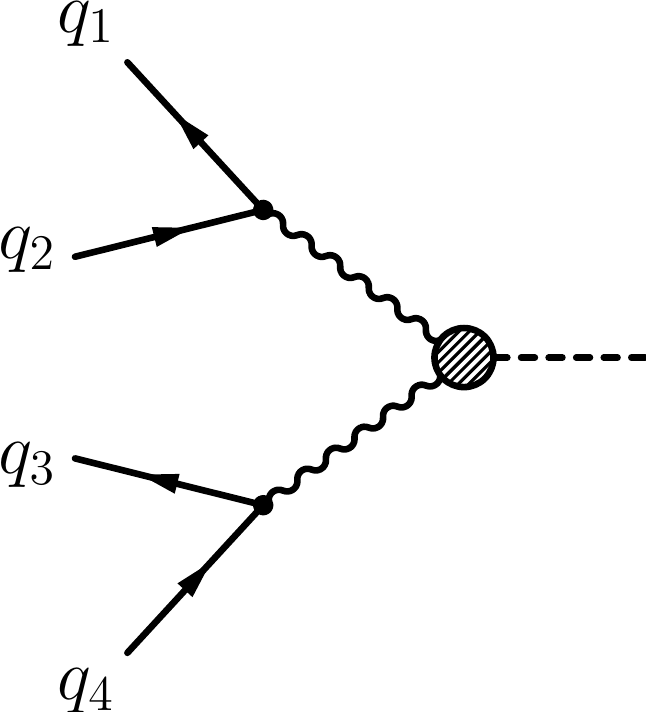}  \hspace{2em}
 \includegraphics[trim = 0 0 0 0, width = 0.15\textwidth]{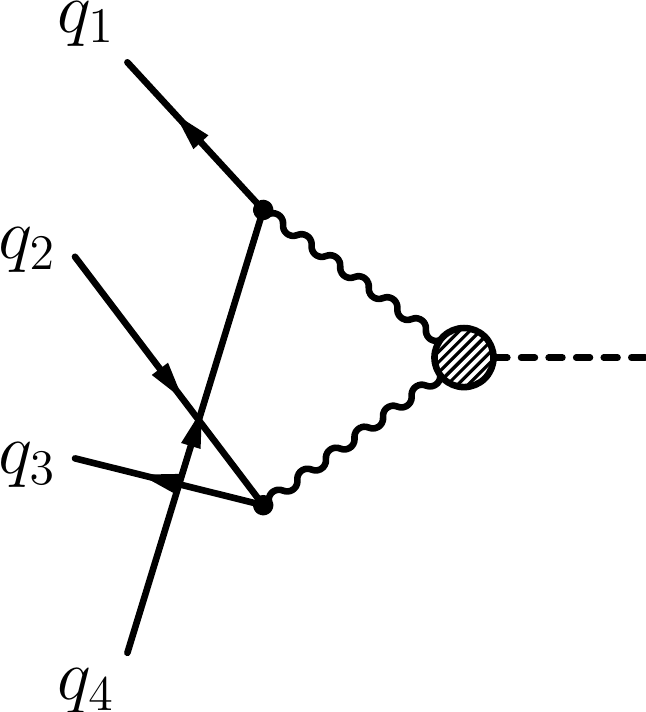}  
 \caption{Possibilities to produce the momenta $q_1$ and $q_3$ of the electrons and $q_2$ and $q_4$ of the positrons for the decay $\pi^0 \rightarrow 2e^+ \, 2e^-$.}
 \label{fig:dd-feyn}
 \end{centering}
\end{figure}

%
%
%
%
For the double-virtual decay of a neutral pion into two dielectrons, the general matrix element is given by
\begin{align}
 &\mathcal{M}_{\pi^0 \rightarrow 2e^+ \, 2e^-} \nn \\ 
 &\hspace{0.5em}=  e^3 \, f_{\rm d.v.}(q^2,k^2) \, \eps_{\mu\nu\alpha\beta}\, \left. \frac{q^\mu \, k^\nu}{q^2 \, k^2} \right|_{q=q_1+q_2, \, k=q_3+q_4} \nn \\
 &\hspace{0.5em}\phantom{= +} \, \cdot  \bar{u}(q_{1}, \lambda_{1}) \, \gamma^\alpha \, v(q_2, \lambda_2) \, \bar{u}(q_{3}, \lambda_{3}) \, \gamma^\beta \, v(q_{4}, \lambda_{4}) \nn \\
 &\hspace{0.5em}\phantom{=}{} - e^3 \, f_{\rm d.v.}(q^2,k^2) \, \eps_{\mu\nu\alpha\beta}\, \left. \frac{q^\mu \, k^\nu}{q^2 \, k^2} \right|_{q=q_1+q_4, \, k=q_2+q_3}  \nn \\
 &\hspace{0.5em}\phantom{= -} \, \cdot \bar{u}(q_{1}, \lambda_{1}) \, \gamma^\alpha \, v(q_4, \lambda_4) \, \bar{u}(q_{3}, \lambda_{3}) \, \gamma^\beta \, v(q_{2}, \lambda_{2}) \label{eq:Mdd}
\end{align}
with the double-virtual form factor $f_{\rm d.v.}(q^2,k^2)$. For the decay into two (identical) dielectrons, the measured momenta $q_1$ and $q_3$ of the electrons and $q_2$ and $q_4$ of the positrons can be produced from the two possibilities shown in Fig. \ref{fig:dd-feyn} yielding the two terms in the matrix element. Note that, since positrons are fermions, the exchange of the two positrons between the two diagrams yields an extra factor of $(-1)$ in the matrix element.

The partial decay width for the decay into two dielectrons is defined as \cite{PDG}
\begin{align}
 \Gamma_{\pi^0 \rightarrow 2e^+ \, 2e^-} = \frac{1}{2!\, 2!} \, \int \te{d} \Phi_4(p;q_1,q_2,q_3,q_4) \,\frac{(2\pi)^4}{2m_{\pi^0}} \, \overline{\left|\mathcal{M} \right|^2} \label{eq:G2d}
\end{align}
integrating over the four-body phase space 
\begin{align}
 \te{d}\Phi_4(p;q_1,q_2,q_3,q_4) = \delta^{(4)}\left( p - \sum_{i=1}^4 q_i \right) \prod_{i=1}^4 \frac{\te{d}^3q_i}{(2\pi)^3 \,2E_i}\,.
\end{align}
The width can be rewritten as a sum, $\Gamma_{\rm \hspace{-0.2em} int.} + \Gamma_{\rm \hspace{-0.1em} dir.}$, consisting of two contributions: An interference contribution $\Gamma_{\rm \hspace{-0.2em} int.}$ with terms depending on all possible four-momentum combinations $q_i +q_j$ generating the momenta of the two virtual photons, and a direct contribution $\Gamma_{\rm \hspace{-0.1em} dir.}$ with terms depending only on two such combinations. The first kind of terms arises from multiplying the two terms in the matrix element \eqref{eq:Mdd} and the second kind from multiplying terms with themselves. 
For the direct contribution, the 12-dimensional phase space integral \eqref{eq:G2d} can be reduced to a two-dimensional integral over the four-momenta of the virtual photons \cite{diplCarla},
\begin{align}
 \Gamma_{\rm \hspace{-0.1em} dir.} =& \,\frac{e^6}{(2\pi)^5 \, 36 \, m_{\pi^0}^3} \,\int \dint q^2 \, \dint k^2 \, \Theta_{\rm k.c.}(q,k) \, \frac{1}{q^2 k^2} \nn \\
 &\hspace{-0.5em} \cdot \sqrt{1 - \,\frac{4m_e^2}{q^2}} \sqrt{1 - \, \frac{4m_e^2}{k^2}} \left(1 + \,\frac{2m_e^2}{q^2} \right) \left(1 + \,\frac{2 m_e^2}{k^2}\right)\nn \\
 &\hspace{-0.5em} \cdot \left[ \frac{1}{4}\left(m_{\pi^0}^2 - (q^2+k^2) \right)^2 - q^2 k^2 \right]^{3/2} \left| f_{\rm d.v.}(q^2,k^2) \right|^2 \nn \\
\end{align}
including the kinematical constraints
\begin{align}
 \Theta_{\rm k.c.}(q,k) \ = \phantom{\cdot}& \Theta\hspace{-0.2em}\left(q^2 -4 m_e^2 \right) \Theta\hspace{-0.2em}\left(k^2- 4m_e^2 \right) \nn \\
 & \cdot \Theta\hspace{-0.2em}\left(m_{\pi^0} - \sqrt{q^2} - \sqrt{k^2} \right).
\end{align}
The interference contribution can only be reduced to a five-dimensional integral \cite{diplCarla},
\begin{align}
 \Gamma_{\rm \hspace{-0.2em} int.} =&\, \frac{e^6}{(2\pi)^6 \, 16 \, m_\pi} \, \int \dint |q_1| \, \dint |q_2| \, \dint |q_3| \, \dint \cos \theta_2 \, \dint \cos\theta_3 \nn \\
 &\hspace{1em} \cdot  \Theta_{\rm k.c.}(q,k) \, \Theta_{\rm k.c.}(q\hspace{0.05em}',k\hspace{0.05em}') \, \Theta(E_4 - m_e) \, \Theta\hspace{-0.1em}\left(1 - \left| \alpha \right| \right)\nn \\
 &\hspace{1em} \cdot \left[ \left(1-\cos^2\theta_2\right) \left(1-\cos^2\theta_3\right) \left(1- \alpha^2\right) \right]^{-1/2}\nn \\
 &\hspace{1em} \cdot \frac{\left|\vec{q}_1\right|^2 \left|\vec{q}_2\right| \left|\vec{q}_3\right|}{E_1 E_2 E_3} \, P_{4l} \, f_{\rm d.v.}(q^2,k^2) \, f^\ast_{\rm d.v.}(q\hspace{0.05em}'^{\,2}, k\hspace{0.05em}'^{\,2}). \nn \\
\end{align}
Here, the momenta of the virtual photons are given by
\begin{align}
 &q = q_1 + q_2, \,\, k = q_3 + q_4, \, \, q\hspace{0.05em}' = q_1 +q_4, \,\, k\hspace{0.05em}' = q_2 +q_3
\end{align}
and $\alpha$ denotes the combination
\begin{align}
 \alpha =&\, \frac{1}{2 \left|\vec{q}_2 \right| \left|\vec{q}_3 \right| \sin\theta_2 \, \sin\theta_3} \left\{ \left(m_\pi^0 - E_1 -E_2 - E_3 \right)^2 \right. \nn \\
 &{}- m_e^2 - \left|\vec{q}_1 \right|^2 - \left|\vec{q}_2 \right|^2 - \left|\vec{q}_3 \right|^2  - 2\left| \vec{q}_1\right| \left| \vec{q}_2 \right| \cos\theta_2 \nn \\
 &{} - 2 \left| \vec{q}_1\right| \left| \vec{q}_3 \right| \cos\theta_3  - 2\left| \vec{q}_2\right| \left| \vec{q}_3 \right| \cos\theta_2  \cos\theta_3 \left. \vphantom{\left(m_{\pi^0} - E_1 -E_2 - E_3 \right)^2} \right\}. 
\end{align}
Furthermore, the phase-space factor is given by
\begin{align}
 P_{4l} =& - \eps^{\mu\nu\alpha\beta} \, k_\mu \,q_\nu \, \eps^{\bar{\mu}\bar{\nu}\bar\alpha\bar\beta} \, k\hspace{0.05em}'_{\bar\mu} \,q\hspace{0.05em}'_{\bar\nu} \nn \\
 &\cdot \sum_{\lambda_i} \ \bar{u}(q_1, \lambda_1) \, \gamma_\alpha \, v(q_2,\lambda_2) \, \bar{u}(q_3, \lambda_3) \, \gamma_\beta \, v(q_4,\lambda_4) \, \nn \\[-1em]
 & \phantom{\sum_{\lambda_i} \ } \,\cdot \bar{v}(q_2,\lambda_2) \, \gamma_{\bar{\beta}} \, u(q_3,\lambda_3) \, \bar{v}(q_4,\lambda_4) \, \gamma_{\bar{\alpha}} \, u(q_1,\lambda_1)\,. \nn \\[-1em]
\end{align}
As the five-fold integral for the interference contribution $\Gamma_{\rm \hspace{-0.2em} int.}$ has to be determined numerically, the full partial decay width is approximated by the direct contribution $\Gamma_{\rm \hspace{-0.1em} dir.}$ in most applications assuming that the interference contribution is very small \cite{Kroll:1955zu}. We will present results for both contributions to study the influence of the interference part. 

The Lagrangians \eqref{eq:LWZW}, \eqref{eq:Lvec} and \eqref{eq:LA} yield the double-virtual form factor\footnote{Note that, of course, $f_{\rm s.v.}(q^2) = f_{\rm d.v.}(q^2,0)$.}
\begin{align}
 f_{\rm d.v.}(q^2,k^2) =&{} - \frac{e\, f_V^2 \, h_A}{6 \, \fpi} \left\{ S_\rho(q^2) S_\omega(k^2) \left(q^2 + k^2 \right) \right. \nn \\
 &{} \left. + (q \leftrightarrow k) \right\} + \,\frac{n \, e}{12 \, \pi^2 \,\fpi}\,. \label{eq:Ff-dv}
\end{align}
Again, this form factor is compared to the VMD prediction.
In order to do so, the form factor is normalized to the double-photon point $(q^2, k^2) =0$, 
\begin{align}
 F_{\rm d.v.}(q^2,k^2) := \frac{f_{\rm d.v.}(q^2,k^2)}{f_{\rm d.v.}(0,0)} \,. \label{eq:Ff-dv-norm}
\end{align}
For an averaged $\rho$/$\omega$ mass $m_V$, the VMD prediction is given by
\begin{align}
 F^{\rm VMD}_{\rm d.v.}(q^2,k^2) &= \frac{1}{2} \left\{ \frac{m_\rho^2}{m_\rho^2 - q^2} \, \frac{m_\omega^2}{m_\omega^2 -k^2} \, + \left(q \leftrightarrow k \right) \right\} \nn \\
 &\approx \frac{m_V^4}{\left(m_V^2 - q^2 \right) \left(m_V^2 - k^2 \right)} . \label{eq:Ff-dv-VMD}
\end{align}
On the other hand, the normalized form factor \eqref{eq:Ff-dv-norm} can be approximated for an averaged $\rho$/$\omega$ mass and $4 \pi^2 f_V^2 h_A / |n| \approx m_V^2$ (cf. discussion after Eq. \eqref{eq:Ff-pi-VMD}) as 
\begin{align}
 F_{\rm d.v.}(q^2,k^2) =&{} \, 1 -\,\frac{2\pi^2  f_V^2  h_A}{n} \left\{ \frac{q^2+k^2}{\left(m_\rho^2 - q^2\right) \left(m_\omega^2 -k^2\right)}  \right. \nn \\[-0.7em]
 &{}  \phantom{-\,\frac{\pi^2  f_V^2  h_A}{n}\ \ \  \left\{ \frac{q^2+k^2}{\left(m_\rho^2 - q^2\right)} \right\}} + \left. \vphantom{\frac{q^2+k^2}{\left(m_\rho^2 - q^2\right)}} \left(q \leftrightarrow k \right) \right\} \nn \\
 \approx& \, \frac{m_V^4 + q^2 k^2}{\left(m_V^2 - q^2 \right) \left(m_V^2 - k^2 \right)} \,.
\end{align}
In contrast to the single-virtual decay \eqref{eq:Ff-sv-approx}, our prediction and the VMD prediction for the double-virtual decay disagree.			

For the form factor \eqref{eq:Ff-dv} and both $h_A = 2.17$ and $h_A = 2.02$ the direct and interference contribution to the partial decay width are given by
\begin{align}
 \Gamma_{\rm \hspace{-0.1em} dir.} &= 2.70 \cdot 10^{-13} \, \te{GeV}, \\
 \Gamma_{\rm \hspace{-0.2em} int. \hspace{0.1em}} &= - 0.02 \cdot 10^{-13} \, \te{GeV}.
\end{align}
The full partial decay width is given by
\begin{align}
 \Gamma_{\pi^0 \rightarrow 2e^+ \, 2e^-} =  \Gamma_{\rm \hspace{-0.1em} dir.} + \Gamma_{\rm \hspace{-0.2em} int.} = 2.68 \cdot 10^{-13} \, \te{GeV} 
\end{align}
in very good agreement with the experimental value \cite{PDG},
\begin{align}
 \Gamma_{\pi^0 \rightarrow 2e^+ \, 2e^-}^{\rm \, exp.} = (2.58 \pm 0.13) \cdot 10^{-13} \, \te{GeV} \,.
\end{align}
Thereby, the deviation of the direct contribution from the full width is less than $1\%$ and confirms the treatment of the interference contribution as negligible. However, this relation is the overall value. The deviation could be quite different in parts of the kinematically allowed region. Additionally, it was shown in \cite{diplCarla} that the interference contribution to the decay of an $\etapr$-meson into two dimuons is more than $10\%$.

For the VMD form factor \eqref{eq:Ff-dv-VMD}, only the normalized double-virtual form factor is accessible. Hence, one can only calculate branching ratios,  
\begin{align}
 \Gamma_{\rm \hspace{-0.1em} dir.}^{\rm \, VMD} \,/\, \Gamma_{\pi^0 \rightarrow 2 \gamma}^{\rm \, exp.} &= 3.46 \cdot 10^{-5}, \\
 \Gamma_{\rm \hspace{-0.2em} int.}^{\rm \, VMD} \,/\,  \Gamma_{\pi^0 \rightarrow 2 \gamma}^{\rm \, exp.} &= -0.02 \cdot 10^{-5}
\end{align}
yielding the full branching ratio 
\begin{align}
 \Gamma_{\pi^0 \rightarrow 2e^+ \, 2e^-}^{\rm \, VMD} \,/\, \Gamma_{\pi^0 \rightarrow 2 \gamma}^{\rm \, exp.} = 3.44 \cdot 10^{-5}\,.
\end{align}
Here, the interference contribution is also less than $1\%$ of the full width. The result is in good agreement with the experimental value, 
\begin{align}
 \Gamma_{\pi^0 \rightarrow 2e^+ \, 2e^-}^{\rm \, exp.} \,/\, \Gamma_{\pi^0 \rightarrow 2 \gamma}^{\rm \, exp.} = (3.38 \pm 0.13) \cdot 10^{-5}\,.
\end{align}
If one normalizes the result for the width in our approach to our result for the decay into two photons \eqref{res:G2g}, one gets the same numbers as for the VMD branching ratio in contrast to the predicted difference in the double-virtual form factors. This is caused by the small available phase space, the predictions for both form factors barely differ from the normalization value $1$.\\
Furthermore, our results are in agreement with results from modified VMD and the hidden gauge model \cite{Petri:2010ea}.

\section{Scattering $e^+ e^- \rightarrow \pi^+ \pi^- \pi^0$} \label{sec:scattering}
\begin{figure}[h]
 \centering
 \includegraphics[width = 0.25\textwidth]{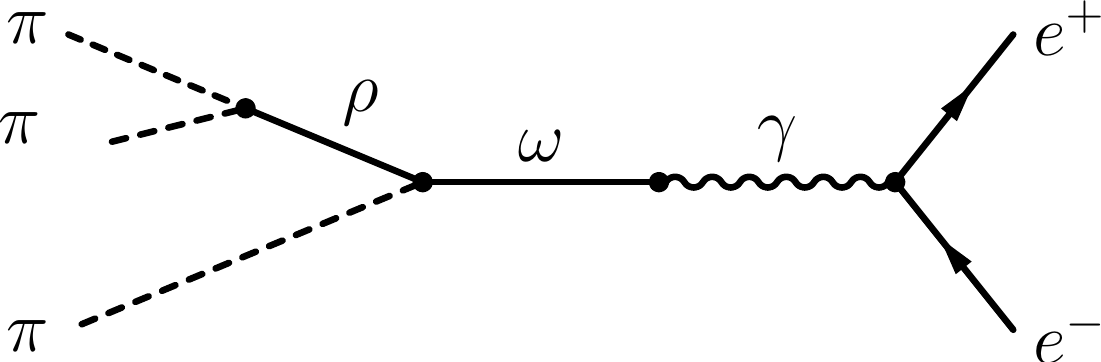} \hspace{2em}
 \includegraphics[width = 0.18\textwidth]{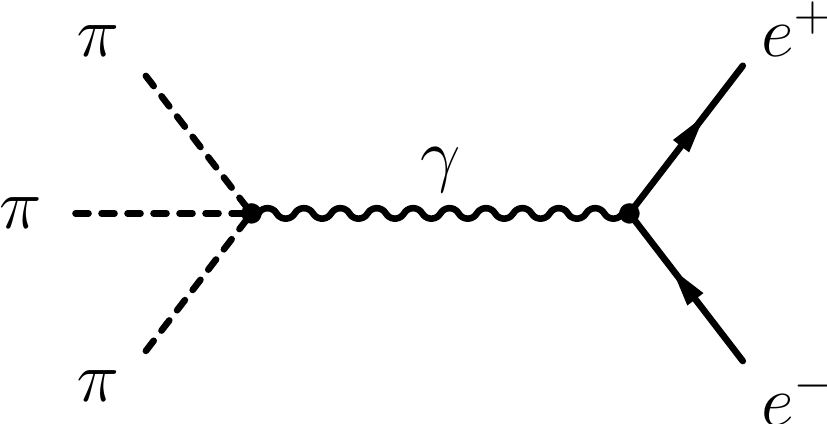} 
 \caption{Scattering $e^+ e^- \rightarrow 3\pi$ via an $\omega$-meson (\textbf{left}) and direct (\textbf{right}).}
 \label{fig:ee3pi}
\end{figure}

The consideration of the decay $\omega \to 3\pi$ in section \ref{sec:omega decays} can be used to describe the scattering $e^+ e^- \to 3\pi$ \cite{Strandberg:2012}: Since the $\omega$-meson couples to a photon ($f_V$ term in \eqref{eq:Lvec}), the scattering reaction can happen via an $\omega$-meson (left-hand side in Fig. \ref{fig:ee3pi}). Additionally, the WZW Lagrangian \eqref{eq:LWZW} describes the direct reaction (right-hand side in Fig. \ref{fig:ee3pi}). We will evaluate the influence of pion-pion rescattering and the WZW contribution on the total cross section. Note that the reactions $e^+ e^- / \omega \to 3\pi$ have also been considered in \cite{Czyz:2005as}.

The general differential cross section for a total reaction energy $\sqrt{s}$ in the center-of-mass system is given by
\begin{align}
 \frac{\dint \sigma_{ee \rightarrow 3\pi}}{\dint m_{12}^2 \, \dint m_{23}^2} = \frac{s +2 m_e^2}{32 \, (2\pi)^3 \, \sqrt{s}^{\,7} \, \sqrt{s - 4m_e^2}} \, P_{3\pi} \left|C_{ee\rightarrow 3\pi} \right|\,
\end{align}
with $P_{3\pi}$ defined in \eqref{eq:P3pi}. For the contribution via an $\omega$-meson described by \eqref{eq:Lvec} and \eqref{eq:LA}, the reduced matrix element equals
\begin{align}
 C_{ee \rightarrow 3\pi}^{\rm vec} = - \, \frac{h_P \, h_A \, f_V^2 \, e^2}{12 \, \fpi^3} \, S_\omega(s) \hspace{-1em}  \sum_{\substack{(i,j) = (1,2),\\ (2,3), (1,3)}} \hspace{-1em}  S_\rho(m_{ij}^2) \left( m_{ij}^2 + s \right). \nn \\[-1.5em] \label{eq:ee3pi vec-ME}
\end{align}
Note that this relation corresponds to \eqref{eq:Matrixelemen omega3pi} but with $m_\omega^2$ replaced by $s$. The two-body variables for the three-pion system $m_{ij}^2 = (p_i + p_j)^2$ satisfy now
\begin{align}
 m_{12}^2 + m_{23}^2 + m_{13}^2 = 2 m_{\pi^+}^2 + m_{\pi^0}^2 + s \,.
\end{align}
To incorporate pion-pion rescattering, the $\rho$-meson propagator has to be replaced in the same manner as done in Eq. \eqref{eq:resc-prop} in section \ref{sec:omega decays}. Here, it is reasonable to use the value for the parameter combination $h_A \,f_V / \fpi$ obtained from the decay $\omega \to 3\pi$ since this is the final channel in the reaction $e^+ e^- \to 3\pi$. Strictly speaking, the formula for the $\omega \rightarrow 3\pi$ width should enter the denominator of the $S_\omega$ propagator in \eqref{eq:ee3pi vec-ME}. Since the $\omega$ is a very narrow state, we use a constant width instead. $90\%$ of the total width of the $\omega$-meson are determined by its decay into three pions and merely about $10\%$ by the decay into a neutral pion and a real photon. Using our value for $h_A$ as obtained from $\omega \rightarrow 3\pi$ we get the correct value for $90\%$ of the total $\omega$ width. If the rest ($\omega \rightarrow \pi^0 \gamma$) is calculated with the value for $h_A \,f_V / \fpi$ obtained  from $\omega \to 3\pi$, it differs about $20\%$ from experiment. Thus, the total width is only changed by about $2\%$ which is less than the accuracy we can achieve. The uncertainties in the reaction $e^+ e^- \to 3 \pi$ will be estimated by varying the vector-meson decay constant $f_V$ in the incoming channel, i.e. $\gamma^{\ast} \to \omega^\ast$, between the values obtained from $\omega \rightarrow l^+ l^-$ and from the pion form factor, $140$ and $150\,\te{MeV}$, respectively. 

The matrix element calculated with the WZW Lagrangian \eqref{eq:LWZW} is given by
\begin{align}
 C_{ee \rightarrow 3\pi}^{\rm WZW} = \frac{n \,e^2}{12  \pi^2 \, \fpi^3}\,
\end{align}
which enlarges to 
\begin{align}
 C_{ee \rightarrow 3\pi}^{\rm WZW} \ \longmapsto \ C_{ee \rightarrow 3\pi}^{\rm WZW} &\left\{ 1 + \sum_{m_{ij}} t_{l=1,I=1}(m_{ij}) \right. \nn \\[-2em]
 &\left. \phantom{\left\{ 1 + \sum_{m_{ij}} \right.} \cdot \left(J(m_{ij}^2) - \te{Re}\,J(\mu^2)\right) \right\}
\end{align}
in the case of rescattering. 

\begin{figure}[h]
 \centering
 \includegraphics[trim = 40 50 0 60, width = 0.5\textwidth]{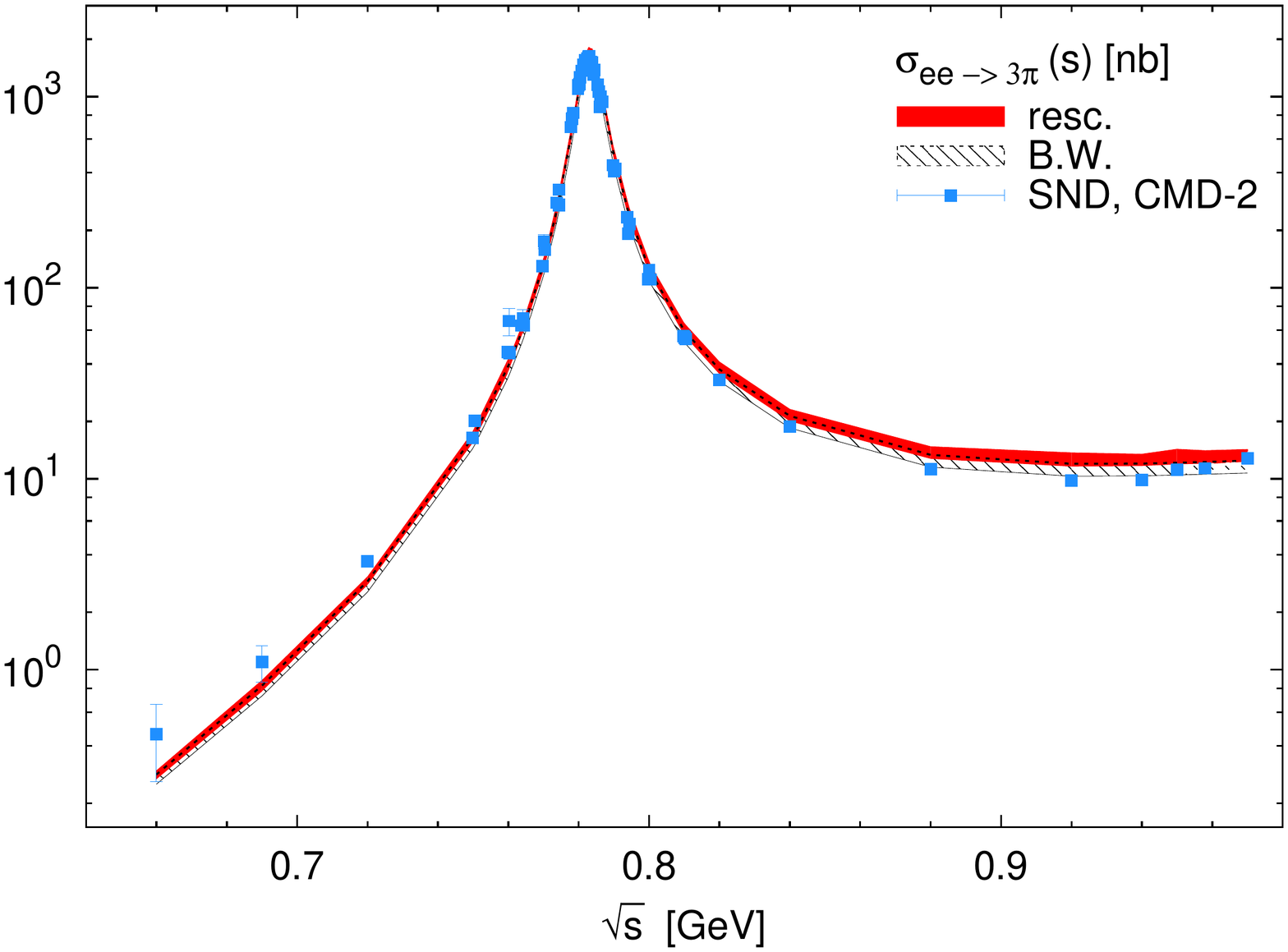}
 \includegraphics[trim = 40 50 0 60, width = 0.5\textwidth]{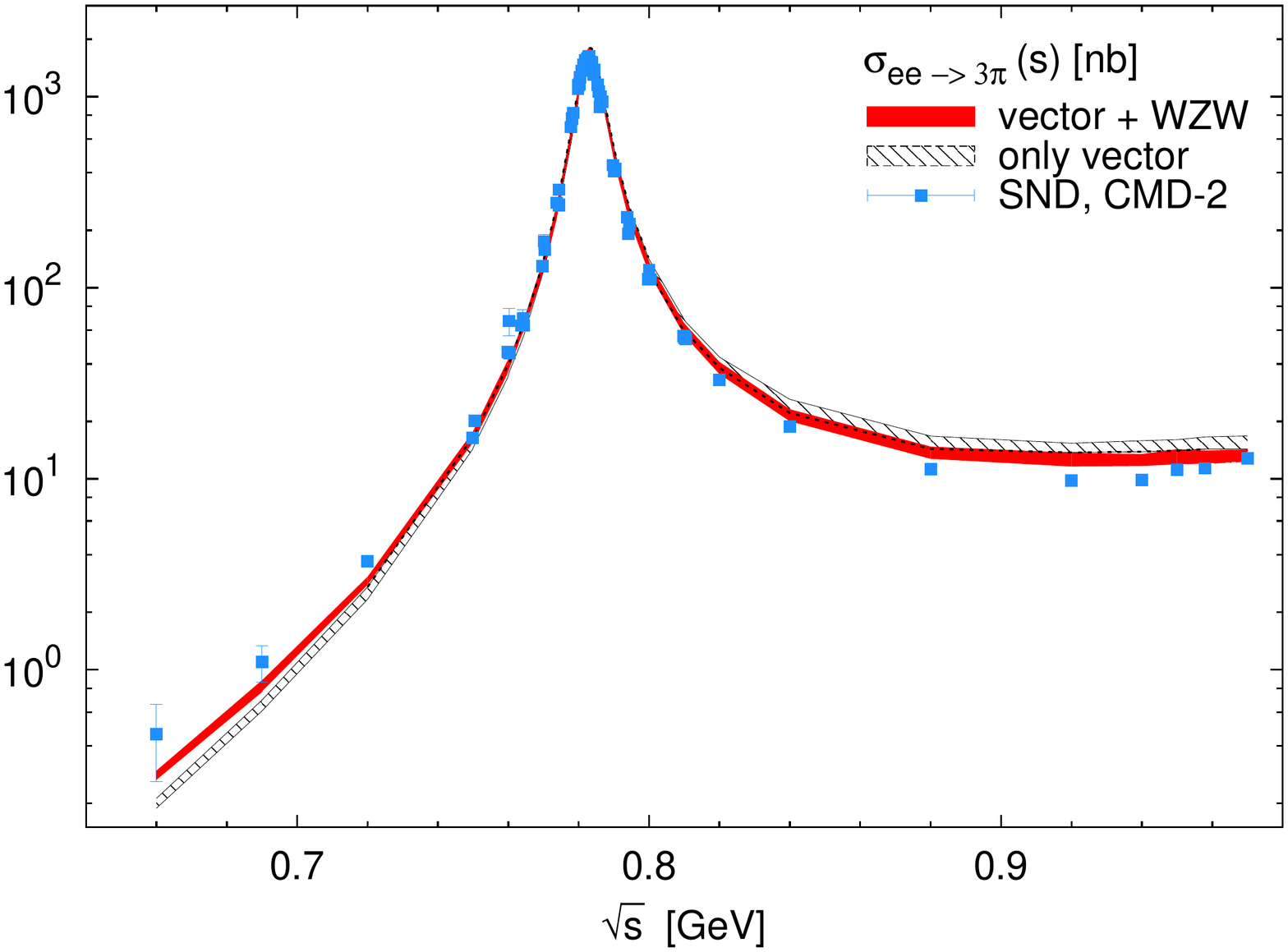}
 \caption{Cross section for the reaction $e^+ e^- \rightarrow 3\pi$ compared to experimental data taken by the SND detector \cite{data1} and by the CMD-2 detector \cite{data2}. \textbf{Top:} The solid red band corresponds to the calculation with rescattering and the dashed black band to the one with a manually added Breit-Wigner width in the $\rho$-meson propagator. \textbf{Bottom:} The solid red line corresponds to the full calculation with WZW and vector contribution and the dashed black line to the one with only the vector contribution. Here, both calculations were done with rescattering.} 
 \label{fig:ee3pi-Resc}
\end{figure}

In Fig. \ref{fig:ee3pi-Resc}, the calculated final cross section formula of the reaction $e^+ e^- \rightarrow 3\pi$ is compared to experimental data taken by the SND detector \cite{data1} and by the CMD-2 detector \cite{data2} for reaction energies $0.66 \, \te{GeV} \leq \sqrt{s} \leq 0.97 \, \te{GeV}$. We do not compare to data above $0.97 \,\te{GeV}$ because there inelastic channels like $e^+ e^- \rightarrow K \bar{K} \rightarrow 3 \pi$ may become important. Additionally, the data shows the importance of the reaction via a virtual $\phi$ meson for higher energies. The decay $\phi \rightarrow 3 \pi$ is not included in the simple vector-meson Lagrangian used in this publication. The error bands in both figures emerge from varying $f_V$ between $140$ and $150 \,\te{MeV}$. For all calculations the error bands and hence the estimated uncertainty are small.

In the top panel of Fig. \ref{fig:ee3pi-Resc}, the cross section calculated with rescattering (solid red band) and using the form \eqref{eq:rho-width full} for the $\rho$-meson propagator where a Breit-Wigner width was added manually (dashed black band) are compared to the experimental data. Both calculations describe the data very well. As expected, the difference between the two calculations is small for energies below the $\rho$-meson mass and the effect of rescattering becomes larger with higher energies.

\begin{figure}[ht]
 \begin{centering}
 \includegraphics[trim = 40 50 0 60, width = 0.5\textwidth]{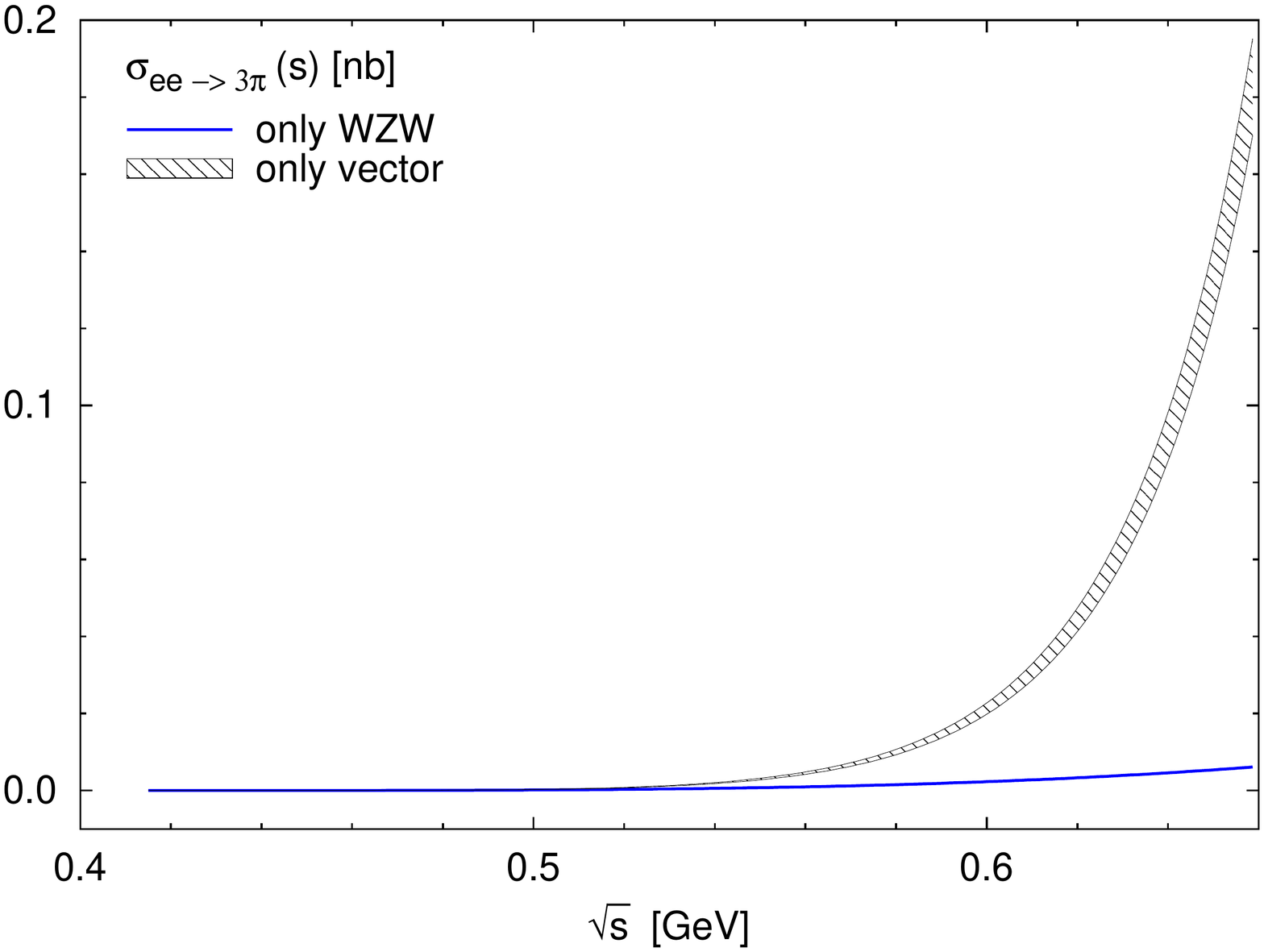}
 \includegraphics[trim = 40 50 0 60, width = 0.5\textwidth]{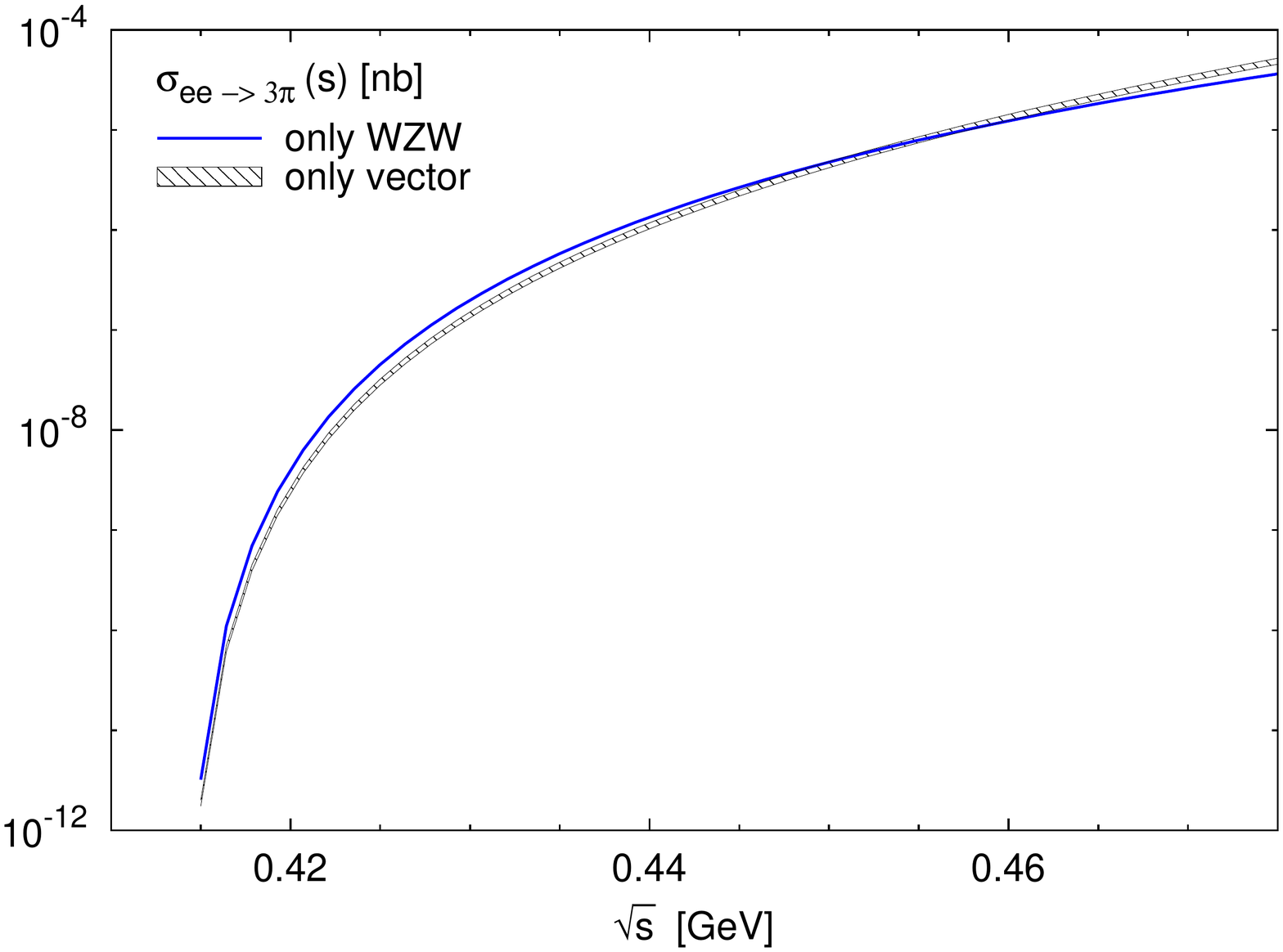}
 \caption{Comparison of the cross section of the reaction $e^+ e^- \rightarrow 3\pi$ using the WZW interaction (solid blue band) and via the $\omega$-meson (dashed black band) for energies near the three-pion threshold. \textbf{Top}: Energy regime $ 3m_\pi \leq \sqrt{s} \leq 0.66 \,\te{GeV}$. \textbf{Bottom:} Energy regime $3m_\pi \leq \sqrt{s} \leq 0.49 \,\te{GeV}$.}
 \label{fig:ee3pi-vecWZW}
 \end{centering}
\end{figure}

Additionally, we compare the calculation with the full matrix element to the one involving only the vector-meson matrix element \eqref{eq:ee3pi vec-ME} (bottom panel of Fig. \ref{fig:ee3pi-Resc}). Thereby, both calculations include rescattering. The WZW term makes a difference at lowest energies and at energies above the $\omega$-meson peak. Incorporating the WZW term improves the calculations in these regions. However, the pure $\chi$PT calculation without the vector meson contribution would not describe the data. For illustration, we also compared calculations with merely the WZW term (solid blue line) and merely the vector-meson contribution (dashed black line) for energies between the three-pion threshold and $0.66 \, \te{GeV}$ in the top panel of Fig. \ref{fig:ee3pi-vecWZW}\footnote{To the best of the authors' knowledge there is no cross-section data available for energies below $0.66 \, \te{GeV}$.}. From about $0.46 \, \te{GeV}$ on, the vector-meson contribution is more important than the WZW contribution. Below that, the WZW contribution is slightly more important (see bottom panel of Fig. \ref{fig:ee3pi-vecWZW}). Obviously, there is no kinematically accessible region where the WZW contribution dominates. In other words, the three-pion threshold is already rather far away from the chiral limit.
\\

\section{Further discussion} \label{sec:fudi}

As demonstrated in the previous sections our Lagrangian provides a good description of a large amount of data using only the three coupling constants $f_V$, $h_P$ and $h_A$ (together with the pion mass, the vector-meson mass and the pion decay constant we have in total six parameters). This raises the question whether there is a deeper foundation
of this Lagrangian which would discriminate it from other phenomenological approaches. In particular, one might ask why the contact terms $V$-$3\pi$ and $V$-$\gamma$-$\pi$ seem to be small/irrelevant when one uses antisymmetric tensor fields to represent the vector mesons. In principle, one can easily imagine that a term
\begin{align}
\varepsilon^{\mu\nu\alpha\beta} \, {\rm tr}(D^\lambda V_{\lambda\mu} \, U_\nu \, U_\alpha \, U_\beta)
\label{eq:smallV3P}
\end{align}
competes with the two-step process of vector-meson exchange $\sim h_A \, h_P$. 
As we have seen, one does not need such a term (\ref{eq:smallV3P}) to describe $\omega \to 3 \pi$ and 
$e^+ e^- \to 3\pi$. The same remark applies to a contact term for $V$-$\gamma$-$\pi$,
\begin{align}
i \, \varepsilon^{\mu\nu\alpha\beta} \, {\rm tr}(\{D^\lambda V_{\lambda\mu},f^+_{\nu\alpha} \} \, U_\beta)  \,.
\label{eq:smallVgP}
\end{align}
It would contribute, e.g., to the pion transition form factor, but apparently it is not needed.

One way to justify the suppression of (\ref{eq:smallV3P}) and (\ref{eq:smallVgP}) has been expressed in 
\cite{Terschlusen:2012xw, Lutz:2008km}. Therein an effective field theory based on the hadrogenesis conjecture has been proposed for light pseudoscalar and vector mesons. In this framework both the light pseudoscalar ($P$) and light vector mesons ($V$) are treated on the same footing yielding the power counting rules 
\begin{align}
 m_V, \, m_P, \, D_\mu \sim p
\end{align}
for a typical momentum $p$.

Actually, this counting scheme allows for a second, numerically less important leading-order term describing the vertex $2V$-$\pi$ in addition to the $h_A$ term \eqref{eq:LA} \cite{Terschlusen:2012xw}. However, the corresponding parameter $b_A$ is less than $10\%$ of the parameter $h_A$. Since the inclusion of the WZW Lagrangian is already a more phenomenological approach, we have neglected the additional contribution to the vertex $2V$-$\pi$ in this publication for reasons of simplicity.

In the power counting scheme proposed in \cite{Terschlusen:2012xw,Lutz:2008km} the following changes relative to $\chi$PT take place: The interaction terms in \eqref{eq:Lvec} contribute to the low-energy constants of $\chi$PT at order $p^4$ \cite{Ecker:1988te,Ecker:1989yg}. In the sector of even intrinsic parity this is next-to-leading order in a pure $\chi$PT counting. In the resonance regime treated in \cite{Terschlusen:2012xw,Lutz:2008km} these terms are promoted to order $p^2$, i.e.\ to leading order. At least qualitatively it makes sense that the importance of resonances grows in the regime where they become active degrees of freedom. Indeed in the scheme of \cite{Terschlusen:2012xw,Lutz:2008km} the contributions in \eqref{eq:Lvec} become as important as the leading-order $\chi$PT Lagrangian given in \eqref{eq:LChPT}. However, in the sector of odd intrinsic parity the hierarchy is not evened out but inverted. The $h_A$ term in \eqref{eq:LA} contributes to the low-energy constants of $\chi$PT at order $p^6$ \cite{Kampf:2011ty}. In that way it provides a correction to the WZW structures \eqref{eq:LWZW}, which are of order $p^4$. In the resonance region the $h_A$ term is promoted to order $p^2$, in line with the promotion of the terms in \eqref{eq:Lvec}. But this implies that the $h_A$ term is now more important than the WZW contribution. In part our results support this picture. For the reaction $e^+ e^- \to 3 \pi$ the vector-meson contributions are always very important and the WZW part might be regarded as subleading. On the other hand, this picture does not apply at all to the pion transition form factor. For instance, in the space-like region (cf.\ Fig. \ref{fig:Ff-pi-ratio}) the WZW term remains important throughout. A clarification of this issue in the framework of \cite{Terschlusen:2012xw,Lutz:2008km} requires a full next-to-leading order $\mathcal{O}(p^4)$ calculation which includes the WZW contribution, but also one-loop terms. Clearly this is beyond the scope of the present work where we follow a more phenomenological approach. 

Treating vector mesons as light is, of course, not without conceptual problems (see also the discussion in \cite{Harada:2003jx}). In particular, the vector-meson masses are close to the scale where loops become important in $\chi$PT. This scale is given by $4\pi \fpi$. On the other hand, the inclusion of resonances as active degrees of freedom requires a serious treatment of unitarity. Resummations of two-particle reducible diagrams are mandatory. Thus, a strict perturbative treatment is anyway not possible and the power counting can (at best) be applied to scattering kernels and not directly to amplitudes. Here the power counting scheme of \cite{Terschlusen:2012xw, Lutz:2008km} has produced very reasonable results in coupled-channel calculations for meson-meson scattering \cite{Danilkin:2011fz} and photon-fusion reactions \cite{Danilkin:2012ua}.

In addition to the previous considerations one might ask whether there are alternative justifications of our phenomenologically successful approach to combine the respective leading-order structures of $\chi$PT, (\ref{eq:LChPT}) and \eqref{eq:LWZW}, with the vector-meson Lagrangians (\ref{eq:Lvec}) and (\ref{eq:LA}) \cite{Leupold:2012qn}. Indeed, if one treats vector mesons as heavy fields \cite{Jenkins:1995vb}, similar in spirit to baryon $\chi$PT, the leading-order terms would be just the $h_A$ term (\ref{eq:LA}) together with the kinetic and mass terms for vector mesons. Thus, the $h_A$ term appears, no matter whether one regards the vector mesons as light or as heavy degrees of freedom. In the latter case, however, there is no room for the single-vector-meson terms of (\ref{eq:Lvec}). The decay of a heavy into light degrees of freedom is hard to treat systematically in a power expansion of momenta (see, however, \cite{Djukanovic:2009zn, Bauer:2011bv}). Therefore, such terms $\sim f_V$, $h_P$ have been excluded by hand in the heavy-vector formalism of \cite{Jenkins:1995vb}. On the other hand, the Lagrangian (\ref{eq:Lvec}) is singled out by the fact that it saturates the low-energy constants of $\chi$PT at next-to-leading order \cite{Gasser:1983yg,Ecker:1988te,Ecker:1989yg}. In this pure $\chi$PT framework the vector mesons are, of course, no active degrees of freedom at all, neither light nor heavy. They are rather treated as to be integrated out, i.e.\ in a sense as static (extremely heavy). Thus, there are reasons for the importance of (\ref{eq:Lvec}) and of (\ref{eq:LA}) also if vector mesons are not treated as light but heavy. The arguments just do not fully fit together (yet?) in a way such that a consistent effective field theory for heavy, but unstable vector mesons can be formulated.

All the presented considerations point to the larger importance of (\ref{eq:Lvec}) and (\ref{eq:LA}) as compared to, e.g., (\ref{eq:smallV3P}) and (\ref{eq:smallVgP}). Whether a valid systematic power counting scheme
emerges from these considerations in one or the other way remains to be seen. 

\section{Summary} \label{sec:Summary}
In the present work, we calculated the decay of an $\omega$-vector meson into three pions, the decays of a neutral pion into two real photons, into a photon and a dielectron and into a double dielectron and the scattering reaction $e^+ e^- \rightarrow 3 \pi$. Therefore, we used on the one hand a simple vector-meson Lagrangian. On the other hand, we used the leading-order $\chi$PT contributions in the sectors of even and odd intrinsic parity.

For the decay $\omega \rightarrow 3\pi$ and the scattering $e^+ e^- \rightarrow 3 \pi$, calculations were performed with and without two-pion rescattering yielding a good description of the available scattering data. Furthermore, the effects of an average isospin-limit pion mass were small.

For the pion decays, we were in good agreement with the experimental data. The predicted form factor for the single-virtual decay is in numeric agreement with the VMD prediction whereas the double-virtual is not. Thereby, the contribution from the WZW term was most important in the kinematically allowed time-like decay region. Data in the space-like region was well described. Similar analyses could be done for $\eta$- and $\eta'$-mesons (see also \cite{diplCarla}).

\section*{Acknowledgments}
This work has been supported by the European Community Research Infrastructure Integrating Activity\\ ``Study of Strongly Interacting Matter'' (HadronPhysics3, Grant Agreement No. 283286) under the Seventh Framework Program of the EU.

\begin{appendix}

\end{appendix}

\bibliographystyle{spphys}       
\bibliography{literature-t2}

\end{document}